\documentclass{amsart}

\usepackage{psfrag}
\usepackage{graphicx}
\usepackage{url}
\usepackage{amsmath,amsfonts}
\usepackage{tikz}

\newcommand\bx{{\mathbf x}}
\newcommand\bp{{\mathbf p}}
\newcommand\bu{{\mathbf u}}

\newcommand\bnabla{{\boldsymbol \nabla}}
\newcommand\pnabla{{\boldsymbol \nabla_\bp}}

\begin{document}

\title{Non-linear Instability of periodic orbits of suspensions of thin fibers in fluids}

\author{Stephen Montgomery-Smith}
\email{stephen@missouri.edu}
\address{Department of Mathematics, University of Missouri, Columbia MO 65211, U.S.A.}

\begin{abstract}
This paper is concerned with difficulties encountered by engineers when they attempt to predict the orientation of fibers in the creation of injection molded plastic parts.  It is known that Jeffery's equation, which was designed to model a single fiber in an infinite fluid, breaks down very badly when applied, with no modifications, to this situation.  In a previous paper, the author described how interactions between the fiber orientation and the viscosity of the suspension might cause instability, which could result in the simple predictions from Jeffery's equation being badly wrong.  In this paper, we give some rigorous proofs of instability using Floquet Theory.  We also show that to obtain exponential instability, it is insufficient to consider only two dimensions, although linear instability is still possible.
\end{abstract}

\maketitle

\section{Introduction}

Predicting the orientation of thin fibers suspended in fluid flows with low Reynolds number finds many industrial applications, for example, in creating parts using injection molded plastics.  One method that has been widely used is to start with the assumption that Jeffery's equation \cite{jeffery:23}, or some variation of it, is a good predictor of the orientation of the fibers.  Jeffery's equation alone has been seen to be a poor predictor of the behavior of fibers in fluids when the volume ratio of fibers is reasonably high.  It is clear that hydrodynamic interactions between the fibers have a very important effect.  One way hydrodynamic effects are modeled is by including diffusion terms to Jeffery's equation \cite{bird:87b}, for example, the Folgar-Tucker equation \cite{folgar:84}.

The author \cite{montgomery-smith:10d} proposed a different mechanism for accounting for hydrodynamic interactions was proposed.  Jeffery's equation was coupled with an anisotropic version of Stokes' equation, where the relationship between the stress and the strain depended upon the orientation of the fibers.  It should be understood that this theory is only offered as one possible explanation for why this might take place.

This paper concentrates on one aspect of how experimental data differs from the pure Jeffery's equation.  Jeffery's equation for fibers with finite aspect ratio, equivalently, where the Jeffery's parameter $|\lambda|<1$, predicts that under a shear flow, the fiber orientation is periodic in time, with period $4\pi/\sqrt{1-\lambda^2}$ divided by shear rate.  (The $4\pi$ is replaced by $2\pi$ if the fiber orientation has certain symmetries.)  We illustrate this periodicity in Figures~\ref{fig2a} and~\ref{fig2a-a2}, where we plot the shear stress and components of the second moment of the distribution of fiber orientations.  This is referred to as Jeffery's `tumbling,' and is not seen in experiments for concentrated suspensions  \cite{anczurowski:67}.

Ideally we would like to perform a full numerical simulation of the Jeffery-Stokes equation under a shear flow, and see if the results conform to experimental results.  In lieu of performing this, we will instead perform a simulation in which the initial perturbations only vary in one of the dimensions.  In this case, we do not expect our results to conform to any experimental results.  But instability can be demonstrated.

The results in this paper are rigorous except that we compute the spectrum of the monodromy matrix numerically.  However in many cases, the spectral radius of the monodromy operator is so large that presumably a careful interval analysis of the calculations would yield the same results.

\section{The coupled Jeffery-Stokes equation}

The version of Jeffery's equation we use solves for $\psi$, the probability distribution of the orientation of fibers, at each point in time and space.  (The original Jeffery's equation was concerned with a single ellipsoidal fiber in an infinite Stokes fluid, and so the version presented in this paper is not seen in the original paper.)  The probability distribution is a function $\psi(\bx,\bp,t)$ of the three variables: space $\bx=(x,y,z)$, time $t$, and orientation $\bp\in S$, where $S = \{\bp = (p_1,p_2,p_3):p_1^2+p_2^2+p_3^2 = 1\}$ is the two dimensional sphere.  Note the isotropic distribution is given by $\psi = 1/4\pi$.  The equations involve the velocity field $\bu = (u_1,u_2,u_3)$, which is a function of space $\bx$ and time $t$.  Associated with the velocity field $\bu$ are the Jacobian matrix $\bnabla\bu = \left[{\partial u_i}/{\partial x_j}\right]_{1\le i,j\le 3}$, the deformation matrix or rate of strain tensor $\mathsf \Gamma = \bnabla\bu + (\bnabla\bu)^T$, and the vorticity matrix $\mathsf \Omega = \bnabla\bu - (\bnabla\bu)^T$.  Jeffery's equation is
\begin{equation}
\label{psi}
\frac{\partial \psi}{\partial t} + \bu \cdot\bnabla\psi = - \tfrac12\pnabla\cdot((\mathsf \Omega\cdot \bp + \lambda(\mathsf \Gamma\cdot\bp - \mathsf \Gamma:\bp\bp\bp))\psi)
\end{equation}
Here $\pnabla$ denotes the gradient on the sphere $S$.

The model that was proposed by the author \cite{montgomery-smith:10d} was to couple Jeffery's equation with how the fiber orientation effects the viscosity of the suspension.  It is stated by Batchelor \cite{batchelor:71} and Shaqfeh et al \cite{shaqfeh:90a} that if the underlying fluid is Newtonian, then the stress-strain relation for slender fibers is
\begin{equation}
\mathsf \sigma = \nu(\beta (\mathbb A:\mathsf \Gamma - \tfrac13 \mathsf I (\mathsf A:\mathsf \Gamma)) + \mathsf \Gamma) - p \mathsf I \\
\end{equation}
Here $\mathsf \sigma$ is the stress tensor, $\mathsf A$ and $\mathbb A$ are respectively the the 2nd and 4th moment tensors
\begin{gather}
\mathsf A = \int_S \bp\bp \, \psi \, d\bp \\
\label{A}
\mathbb A = \int_S \bp\bp\bp\bp \, \psi \, d\bp
\end{gather}
$\nu$ is the Newtonian viscosity that the underlying fluid would have if the fibers were absent (without loss of generality we set $\nu = 1$),  $p$ is the pressure, and $\beta$ is a dimensionless quantity that is related to the volume fraction of the fibers in the fluid.  The quantity $\beta$ represents the extent to which fibers act as `stiffeners' to the fluid motion.  The paper by Sepher et al \cite{sepehr:04} suggests that the order of magnitude of $\beta$ could easily be as large as 50 or 100.

We assume that the velocity field obeys the following incompressible Stokes' equation:
\begin{gather}
\label{stokes}
\bnabla \cdot \mathsf \sigma = 0 \\
\label{incompressible}
\bnabla \cdot \bu = 0
\end{gather}
Since the fluid is incompressible, the pressure $p$ is obtained implicitly, and hence without changing any of the results, we can replace $\frac13 \beta \mathsf A:\Gamma + p$ by a single scalar $q$, so that the stress-strain equation becomes
\begin{equation}
\label{stress}
\mathsf \sigma = \beta \mathbb A:\mathsf \Gamma + \mathsf \Gamma - q \mathsf I \\
\end{equation}

It is known \cite{lipscomb:88,dinh:84,szeri:96,montgomery:10b} that if $\psi$ was ever isotropic at some time in the past, then the solution to equation~\eqref{psi} is
\begin{equation}
\label{psi-B}
\psi(\bp) = \psi_{\mathsf B}(\bp) = \frac1{4\pi (\mathsf B:\bp\bp)^{3/2}}
\end{equation}
where $\mathsf B$ is a symmetric positive definite matrix with determinant one satisfying
\begin{equation}
\label{B-full}
\frac{\partial \mathsf B}{\partial t} + \bu\cdot\bnabla \mathsf B = - \tfrac12 \mathsf B \cdot(\mathsf \Omega+\lambda \mathsf \Gamma) - \tfrac12 (- \mathsf \Omega+\lambda \mathsf \Gamma) \cdot \mathsf B
\end{equation}
Thus from now we will always assume that equation~\eqref{psi-B} is satisfied.  It can be shown that $\mathsf B$ can achieve any positive definite matrix with determinant one, starting from initial data $\mathsf B = \mathsf I$.  Hence any positive definite matrix with determinant one can be used as initial data for $\mathsf B$.  (This quantity $\mathsf B$ is a kind of left Cauchy-Green, or Finger, deformation tensor, mediated by the value of $\lambda$.)

The 4th order moment tensor $\mathbb A$ can be calculated directly from $\mathsf B$ using elliptic integrals \cite{montgomery:10b,verleye:93,verweyst:98}
\begin{equation}
\label{A from B}
\mathbb A = \mathbb A(\mathsf B) = \tfrac34 \int_0^\infty \frac{s\,\mathcal S((\mathsf B+s \mathsf I)^{-1}\otimes(\mathsf B+s \mathsf I)^{-1}) \, ds}{\sqrt{\text{det}(\mathsf B+s \mathsf I)}}
\end{equation}
where $\mathcal S$ is the symmetrization of a tensor, that is, if $\mathbb B$ is a rank $n$ tensor, then $\mathcal S(\mathbb B)_{i_1\dots i_n}$ is the average of $\mathbb B_{j_1\dots j_n}$ over all permutations $(j_1,\dots,j_n)$ of $(i_1,\dots,i_n)$.

\section{Shear flow}

We will model a fluid between two infinite plates, one at $y=0$ that is stationary, and one at $y=W$, which is moving in the direction of the $x$-axis at velocity $V$.  Thus the average shear strain rate is $V/W$.  This gives boundary conditions
\begin{equation}
\label{boundary}
\bu(x,0,z) = (0,0,0), \quad \bu(x,W,z) = (V,0,0)
\end{equation}
We also suppose that the pressure gradient in the $x$ and $z$ directions is zero.

In order to prove non-linear instability, we need only show that a certain class of solutions is unstable under perturbations.  Therefore, from now on we will restrict to the case that $\psi$, and hence $\mathsf B$, as a function of $\bx$ depends only upon $y$.  Since the pressure gradient in the $x$ and $z$ directions is zero, it follows that $\bnabla\bu$, $p$ and $q$ also depend only on $y$.  A simple argument shows that if these assumptions are true at $t = 0$, they remain true for $t>0$.

We will reduce our class of solutions even further, and assume that there is an integer $n$ such that $\psi$ is constant on any short interval $W(i-1)/n < y < Wi/n$ ($1 \le i \le n$).  Again, a simple argument shows that if these assumptions are true at $t = 0$, they remain true $t>0$.  And furthermore, it allows us to replace integrals by sums, with the spacial mesh giving exact solutions.  (That is, while it might appear that we are only approximating integrals with Riemann sums, for our class of solutions the Riemann sum is exactly the same as the integral.)

From now on, we will use the following terminology.
\begin{enumerate}
\item The fiber orientation will be called \emph{uniform} if $\psi$ does not depend upon $\bx$.  That is, all points in the fluid have exactly the same fiber orientation.  In particular, $\bnabla \psi = 0$.
\item The fiber orientation will be called \emph{isotropic} if $\psi = 1/4\pi$, that is, the fibers have equal probability of lying in any direction.
\item The fiber orientation will be called \emph{two dimensional} if (a) $\psi$ does not depend upon $z$, the third component of $\bx$, and (b) $\psi(p_1,p_2,p_3) = \psi(p_1,p_2,-p_3)$, that is, the fiber distribution is symmetric about the $xy$-plane.
\item The fiber orientation will be called \emph{unrestricted} if it isn't necessarily two dimensional.
\end{enumerate}
The reason we call the third situation two-dimensional is that the fiber induces a flow that has no $u_3$ component, and so that fiber orientation and the flow together are symmetric about the $xy$ plane.

Note that $\psi_{\mathsf B}$ is uniform if and only if $\mathbf B$ does not depend upon $\bx$; $\psi$ is isotropic if and only if $\mathsf B = \mathsf I$; and $\psi$ is two-dimensional if and only $\mathsf B$ does not depend upon $z$ and
\begin{equation}
\mathsf B = \left[\begin{matrix}\
\mathsf B_{11} & \mathsf B_{12} & 0 \\
\mathsf B_{12} & \mathsf B_{22} & 0 \\
0 & 0 & \mathsf B_{33}
\end{matrix}\right]
\end{equation}

\section{A heuristic argument for why `tumbling' might not take place}

To give some physical intuition, let us first describe how small perturbations could have a profound effect on breaking up the periodicity in time of the orientation in the case when $\lambda$ is close but not equal to $1$.  In this case, Jeffery's equation predicts that the motion of a single fiber in a simple shear is periodic.  The fiber spends most of its time in near alignment with the direction of the shear, and then after a fixed amount of time quickly flips almost $180^\circ$ so that it is again in near alignment.

A reasonable analog is to consider a large number of walkers on a circular track, where a short portion of the track is made of quicksand.  We assume that it takes each walker exactly five minutes to complete the part of the track that is not quicksand, and exactly fifty-five minutes to complete the part of the track that is quicksand.  Let us suppose that initially there are many walkers spaced equally around the track.

Then at any random time, most of the walkers will be seen to be in the quicksand.  However, every hour, and only for a short amount of time, all the walkers will suddenly and seemingly miraculously be equally spaced around the track.  And this is in essence what the Jeffery's equation predicts if the initial orientation is isotropic, and $\lambda$ is close but not equal to $1$.  Most of the time the fibers will be mostly aligned, but every so often, with a period predicted precisely by Jeffery's equation, the fibers will momentarily be isotropic.  This is what we refer to as Jeffery's `tumbling.'

However it is clear that this periodic behavior is rather delicate.  For example, suppose that after the first half hour that there is a small earthquake.  At this time most of the walkers will be struggling through the quicksand.  But the earthquake throws some of them a little bit ahead, and some of them are a little bit behind.  Suddenly the delicate timing is lost, and we will lose this periodic behavior where every hour the walkers are equally spaced.

In the same way, if we start with isotropic data and apply Jeffery's equation with $\lambda$ close but not equal to one, and then while the fibers are highly aligned introduce a small perturbation to the fiber distribution, then it is reasonable to suppose that afterwords the `tumbling' effect will no longer be observed.

We should add that there are other possible reasons why this `tumbling' might not be seen.  For example, one reasonable suggestion is that the fibers all have slightly different aspect ratios.

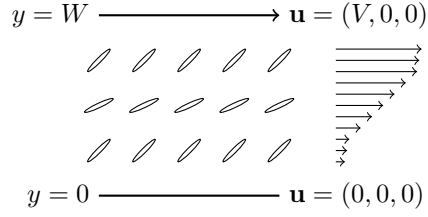
\begin{figure}
\begin{center}
\begin{tikzpicture}\begin{scope}[scale=0.3]
\draw[thick,->] (-4,4)--(4,4);
\draw(-4,4) node[left]{$y=W$};
\draw(4,4) node[right]{$\bu=(V,0,0)$};
\draw(-4,-4) node[left]{$y=0$};
\draw(4,-4) node[right]{$\bu=(0,0,0)$};
\foreach \x in {-4,-2,0,2,4} {
  \draw (0,0) [xshift=\x cm,yshift=2cm,rotate=45] ellipse (0.7 and 0.1);
  \draw (0,0) [xshift=\x cm,yshift=0cm,rotate=25] ellipse (0.7 and 0.1);
  \draw (0,0) [xshift=\x cm,yshift=-2cm,rotate=45] ellipse (0.7 and 0.1);
}
\foreach \x in {6.5} {
  \draw [xshift=\x cm,->] (0,2.5) -- (3.8,2.5);
  \draw [xshift=\x cm,->] (0,2) -- (3.7,2);
  \draw [xshift=\x cm,->] (0,1.5) -- (3.6,1.5);
  \draw [xshift=\x cm,->] (0,1) -- (3.1,1);
  \draw [xshift=\x cm,->] (0,0.5) -- (2.6,0.5);
  \draw [xshift=\x cm,->] (0,0) -- (2.1,0);
  \draw [xshift=\x cm,->] (0,-0.5) -- (1.6,-0.5);
  \draw [xshift=\x cm,->] (0,-1) -- (1.1,-1);
  \draw [xshift=\x cm,->] (0,-1.5) -- (0.6,-1.5);
  \draw [xshift=\x cm,->] (0,-2) -- (0.5,-2);
  \draw [xshift=\x cm,->] (0,-2.5) -- (0.4,-2.5);
}
\draw[thick,-] (4,-4)--(-4,-4);

\end{scope}\end{tikzpicture}
\caption{Shear flow applied to a non-uniform fiber orientation distribution.  (Figure taken from \cite{montgomery-smith:10d}.)}
\label{shear-banding}
\end{center}
\end{figure}

\section{The solution to the simple one-dimensional example}

We show how to compute $\partial \mathsf B/\partial t$ from $\mathsf B$.  First, we obtain $\mathbb A$ using equation~\eqref{A from B}.  Now we show how to compute $\bnabla \bu$.

Equation~\eqref{incompressible} tells us that $u_2$ is constant, and the boundary conditions~\eqref{boundary} tells us that $u_2 = 0$.  Equations~\eqref{stokes} and~\eqref{stress} become
\begin{gather}
\label{stokes1}
(1 + 2\beta \mathbb A_{1122}) \frac{\partial u_1}{\partial y} + 2\beta \mathbb A_{1223} \frac{\partial u_3}{\partial y} = \Sigma_1 \\
\label{stokes2}
2 \beta \mathbb A_{1222} \frac{\partial u_1}{\partial y} + 2 \beta \mathbb A_{2223} \frac{\partial u_3}{\partial y}  - q= \Sigma_2 \\
\label{stokes3}
2\beta \mathbb A_{1223} \frac{\partial u_1}{\partial y} + (1 + 2\beta \mathbb A_{2233}) \frac{\partial u_3}{\partial y} = \Sigma_3
\end{gather}
where $\Sigma_1$, $\Sigma_2$, and $\Sigma_3$ are constant with respect to $y$, but are allowed to depend upon $t$.  We disregard equation~\eqref{stokes2}, as it tells us the pressure, which is information we do not need.

Setting
\begin{equation}
\mathsf N = \mathsf N(y) = \left[\begin{matrix}
1 + 2\beta \mathbb A_{1122} &
2\beta \mathbb A_{1223} \\
2\beta \mathbb A_{1223} &
1 + 2\beta \mathbb A_{2233}
\end{matrix}\right]
\end{equation}
equations~\eqref{stokes1} and~\eqref{stokes3} become
\begin{equation}
\left[\begin{matrix}
\frac{\partial u_1}{\partial y} \\ \frac{\partial u_3}{\partial y}
\end{matrix}\right]
= \mathsf N^{-1} \cdot
\left[\begin{matrix}
\Sigma_1 \\ \Sigma_3
\end{matrix}\right]
\end{equation}
and integrating with respect to $y$ from $0$ to $W$, we obtain
\begin{equation}
\label{V-Sigma}
\left[\begin{matrix}V\\0
\end{matrix}\right]
 =
\left( \int_0^W \mathsf N^{-1}(\eta) \, d\eta\right) \cdot
\left[\begin{matrix}
\Sigma_1 \\ \Sigma_3
\end{matrix}\right]
\end{equation}
that is,
\begin{equation}
\label{Sigma1-V}
\left[\begin{matrix}
\Sigma_1 \\ \Sigma_3
\end{matrix}\right]
 =
\left( \int_0^W \mathsf N^{-1}(\eta) \, d\eta\right)^{-1} \cdot
\left[\begin{matrix}V\\0
\end{matrix}\right]
\end{equation}
and so,
\begin{equation}
\label{du1,du3}
\left[\begin{matrix}
\frac{\partial u_1}{\partial y} \\ \frac{\partial u_3}{\partial y}
\end{matrix}\right]
= \mathsf N^{-1} \cdot
\left( \int_0^W \mathsf N^{-1}(\eta) \, d\eta\right)^{-1} \cdot
\left[\begin{matrix}V\\0
\end{matrix}\right]
\end{equation}
The quantity $\Sigma_1$ is the \emph{shear stress}, that is, the amount of force per unit area of plate exerted in the $x$-direction needed to maintain the constant shear rate $V/W$.  Note that if $\psi$ is uniform, then $\Sigma_1 = V/W$, and $\Sigma_3 = 0$.

Now that we have a formula for $\bnabla \bu$, we can compute $\partial \mathsf B/\partial t$ using equation~\eqref{B-full}.  Note that the calculations are simplified since it is easily shown that $\bu\cdot\bnabla \mathsf B = 0$.

\section{Uniform solutions are Periodic}

We will give explicit solutions when the fiber orientation is uniform.  We will show that the solution is $t_p$-periodic, where
\begin{equation}
\label{t_p}
t_p = \frac{4\pi W}{V \sqrt{1-\lambda^2}}
\end{equation}
and $t_p/2$-periodic if the fiber orientation is two dimensional.

If $\mathsf B$ represents a uniform distribution at time $t = 0$, then it can be shown that $\mathsf B$ represents a uniform distribution, and $\mathsf N$ does not depend upon $y$, for all $t > 0$, and
\begin{equation}
\left[\begin{matrix}
\frac{\partial u_1}{\partial y} \\ \frac{\partial u_3}{\partial y}
\end{matrix}\right]
=\left[\begin{matrix}V/W\\0
\end{matrix}\right]
\end{equation}
Hence
\begin{equation}
\tfrac12 (\mathsf \Omega + \lambda \mathsf \Gamma) = \frac VW \mathsf D
\end{equation}
where
\begin{equation}
\mathsf D = \frac12 \left[\begin{matrix} 0 & \lambda+1 & 0\\ \lambda-1 & 0 & 0\\0&0&0\end{matrix}\right]
\end{equation}
that is
\begin{equation}
\label{dB/dt}
\frac{\partial \mathsf B}{\partial t} = - \frac VW (\mathsf B \cdot \mathsf D + \mathsf D^T \cdot \mathsf B)
\end{equation}
It may be seen by substitution that the solution is
\begin{equation}
\mathsf B = e^{-t\frac VW \mathsf D^T}  \cdot \mathsf B(0) \cdot e^{-t\frac VW \mathsf D}
\end{equation}
Now the eigenvalues of $\mathsf D$ are $\pm \frac i2 \sqrt{1-\lambda^2}$, and hence $\mathsf B$ is $t_p$-periodic.  Furthermore, $e^{-\frac12 t_p \mathsf D} = \text{diag}(-1,-1,1)$, and this commutes with $\mathsf B$ if it represents a two-dimensional fiber orientation.

\section{Analysis of growth of perturbations using linearization}

Now let $\mathsf B_u$ be a uniform, and hence periodic, solution.  Then we can apply the theory of Floquet multipliers \cite[Theorem 2.88]{chicone:06}.  We define the Poincar\'e map
\begin{equation}
P((\mathsf B(0,y))_{0 \le y \le W}) = (\mathsf B(t_p,y))_{0 \le y \le W}
\end{equation}
The derivative of $P$ around $\mathsf B = \mathsf B_u$ is called the monodromy map.  The idea is to compute the eigenvalues of the derivative .  In particular, if we assume that $\mathsf B(y)$ is constant on short intervals $W(i-1)/n < y < Wi/n$, then the Poincar\'e map is a map on a finite dimensional space, and the Floquet theory applies with complete rigor.  That is, the differential equation is non-linearly unstable if the monodromy map has any eigenvalue whose absolute value is larger than 1.

In prior work \cite{montgomery-smith:10d} we have shown the algorithm for computing the eigenvalues of the monodromy map.  They are the eigenvalues of $\mathcal L(\mathbf e_2,t_p)$ as defined in equations~(4.1) and~(8.3) of \cite{montgomery-smith:10d}.  Note there is a sign error in the algorithm \cite{montgomery-smith:10d-corrigendum}: equation~(6.10) should read
\begin{equation}
\tilde{\mathsf\Omega} = i(\hat{\mathbf u} \boldsymbol\kappa - \boldsymbol\kappa \hat{\mathbf u})
\end{equation}
(Note that this equation can only be understood in the context of \cite{montgomery-smith:10d}.)
This sign error makes a huge difference to the numerical results we now give.

Various random values of $\mathsf B_u(0)$ were tried, with the only restriction being that $\mathsf B_u$ be positive definite with determinant one.  Many of them give a value of the largest absolute value of the eigenvalues that is quite small.  But the following matrix gives this value at about $23000$ if $\lambda = 0.98$ and $\beta=10$.  Larger values like $\beta=50$ seemed to cause the program to freeze, suggesting the eigenvalues become very large.
\begin{equation}
\label{the B_0}
\mathsf B_u(0) = 
\left[
\begin{matrix}
 10.7439 & 2.11991 & 3.22009 \\
 2.11991 & 2.93079 & 5.16061 \\
 3.22009 & 5.16061 & 9.15252 \\
\end{matrix}
\right]
\end{equation}
If we are only interested in two-dimensional fiber orientations, then we should consider $\mathsf B_u(0)$ satisfying $\mathsf B_{13}(0) = \mathsf B_{23}(0) = 0$, and consider only the top left $3 \times 3$ submatrix of $\mathcal L(\mathbf e_2,t_p)$.  In that case we find that up to numerical precision, the eigenvalues are bounded by one.  In the next section, we will prove this by showing that the perturbations grow at most linearly.

\section{The simple one-dimensional example when the fiber orientation is two-dimensional}

In this section we will show that if the fiber orientation is two-dimensional, 
then two-dimensional perturbations do grow, but only grow at a linear rate.  Specifically, we shall show that there is a function $\tau(t,y)$, which we will call the \emph{local time}, such that
\begin{gather}
\label{B from tau}
\mathsf B = e^{-\tau \mathsf D^T}  \cdot \mathsf B(0) \cdot e^{-\tau \mathsf D} \\
\label{tau t}
\tau(t,y) = \kappa(y) t + O(1) \text{ as $t \to \infty$}
\end{gather}
where $\kappa(y)$ depends upon $y$, and in general is non-constant.  Note that $\mathsf B$ is periodic in $\tau$ with period
\begin{equation}
\tau_p = \frac{2\pi}{\sqrt{1-\lambda^2}}
\end{equation}
For two-dimensional fiber orientations we have that $\mathbb A_{1223} = 0$.  Hence $\mathsf N(y)$ is a diagonal matrix, and it becomes easy to compute $\mathsf N^{-1}(y)$.  So equation~\eqref{du1,du3} becomes $\partial u_3/\partial y = 0$ and
\begin{equation}
\label{V-Sigma_1}
\frac{\partial u_1}{\partial y} = \left( \int_0^W \frac{dy}{1 + 2 \beta \mathbb A_{1122}} \right)^{-1} \frac{V}{1 + 2 \beta \mathbb A_{1122}}
\end{equation}
This formula can be written as follows.  Define
\begin{gather}
\label{average shear stress}
\text{Average shear stress} = \Sigma_1 = V \left(\int_0^W \frac{dy}{1 + 2 \beta \mathbb A_{1122}}\right)^{-1} \\
\label{average shear viscosity}
\text{Average shear viscosity} = N = \left( \frac1 W \int_0^W \frac{dy}{1 + 2 \beta \mathbb A_{1122}}\right)^{-1} \\
\text{Average shear strain rate} = R = \frac VW
\end{gather}
then equation~\eqref{V-Sigma_1} can be restated as
\begin{equation}
\Sigma_1 = N R
\end{equation}
Note that the average shear viscosity is the harmonic mean of the shear viscosities $(1+2 \beta \mathbb A_{1122})$.  (Note the units in Equations~\eqref{average shear stress} and~\eqref{average shear viscosity} are correct, because we stated at the beginning of the paper that without loss of generality we have $\nu = 1$.)

The force per unit area pushing the top and bottom plates in opposite directions is $\Sigma_1$, and since work done is the integral of force with respect to distance, it follows that the energy expended per unit area of plate after time $t$ is given by
\begin{equation}
\int_0^t \Sigma_1 V \, dt
\end{equation}
We make this quantity dimensionless by dividing by $V$, to give a quantity we call the \emph{normalized energy}
\begin{equation}
E = \int_0^t \Sigma_1 \, dt
\end{equation}
Note that $E$ is a strictly increasing function of $t$, and $E=0$ when $t=0$.  Next, we define $\tau$ as the solution to the differential equation
\begin{gather}
\text{$\tau = 0$ when $E = 0$} \\
\frac{\partial \tau}{\partial E} = \frac1{1+2\beta\mathbb A_{1122}}
\end{gather}
Now
\begin{equation}
\tfrac12 (\mathsf \Omega + \lambda \mathsf \Gamma) = \frac{\partial \tau}{\partial t} \mathsf D
\end{equation}
Hence
\begin{equation}
\label{dB/dtau}
\frac{\partial \mathsf B}{\partial \tau} = - (\mathsf B \cdot \mathsf D + \mathsf D^T \cdot \mathsf B)
\end{equation}
and hence equation~\eqref{B from tau} follows.

Next, it can be seen that $\partial E / \partial \tau$ and $\mathbb A = \mathbb A(\mathsf B)$ are periodic in $E$ with period $E_p$, where
\begin{equation}
E_p = \int_0^{\tau_p} (1+2\beta\mathbb A_{1122}(\mathsf B(\tau))) \, d\tau
\end{equation}
Notice that $E_p$ is a function only of $\mathsf B$ at $t=0$.  Numerical calculations show that $E_p$ depends on $\mathsf B(0)$ in a non-trivial manner, that is, different $\mathsf B$ at $t=0$ will, in general, give rise to different values of $E_p$.  We show example results of calculations for $\lambda = 0.98$ and $\beta = 50$ in table~\ref{ran-mat}.  (Note that we only specify $\mathsf B_{11}$, $\mathsf B_{12}$, and $\mathsf B_{22}$.  The other entries are implied by $\mathsf B$ representing a two-dimensional fiber orientation, and $\det(\mathsf B) = 1$.  The units of $E_p$ are energy per unit area, remembering that we have set $\nu = 1$.  The entries of $\mathsf B$ are unitless.)

\begin{table}
\begin{tabular}{|c|c|}
\hline
$\mathsf B_{11}$, $\mathsf B_{12}$, $\mathsf B_{22}$
&
$E_p$ \\
\hline
1.32539, -2.51612, 5.29462
&
124.374 \\
1.8544, 1.57688, 1.5526
&
140.326 \\
18.1277, 10.6832, 9.56442
&
43.5925 \\
3.23236, 1.66221, 0.899942
&
155.327 \\
5.3305, -4.8482, 5.58189
&
76.7804 \\
0.768847, -1.6146, 4.90368
&
101.545 \\
49.6141, -24.1436, 11.7672
&
151.148 \\
1.04157, 0.660037, 8.18445
&
53.0747 \\
1.62661, 1.48431, 1.60716
&
138.293 \\
14.632, 9.25098, 5.85208
&
161.459 \\
\hline
\end{tabular}

\

\caption{Examples of randomly created matrices, and the associated value of $E_p$.}
\label{ran-mat}
\end{table}

Periodicity implies that
\begin{equation}
\tau = (\tau_p / E_p) E + O(1) \text{ as $E \to \infty$}
\end{equation}
Also, since $1 + 2 \beta \mathbb A_{1122}$ is a periodic function of $E$ with period $E_p$, we have
\begin{equation}
\int_0^E \frac{dE'}{1 + 2 \beta \mathbb A_{1122}(\mathsf B)}
= \frac E{E_p} \int_0^{E_p} \frac{dE'}{1 + 2 \beta \mathbb A_{1122}(\mathsf B)} + O(1) \text{ as $E \to \infty$}
\end{equation}
Therefore
\begin{equation}
t = \int_0^E \frac1V \int_0^W \frac{dy \, dE'}{1 + 2 \beta \mathbb A_{1122}(\mathsf B)} = \kappa_1 E + O(1) \text{ as $E \to \infty$}
\end{equation}
where
\begin{equation}
\kappa_1
= \int_0^W \frac 1{E_p V} \int_0^{E_p} \frac{dE' \, dy}{1 + 2 \beta \mathbb A_{1122}(\mathsf B)}
\end{equation}
Thus it can be shown that equation~\eqref{tau t} holds with $\kappa = E_p/\tau_p \kappa_1$.

\section{Numerical simulations that illustrate instability}

We ran numerical simulations with $\beta=50$, $V=W=1$, and $\lambda=0.98$, and $n = 100$.  The ODE is solved using a standard ODE solver, in our case the Runge-Kutta method of order 4 with step size $0.01$, noting that varying the step size to $0.02$ made negligible difference to the solutions.

\begin{figure}
\begin{center}
\includegraphics[scale=0.4]{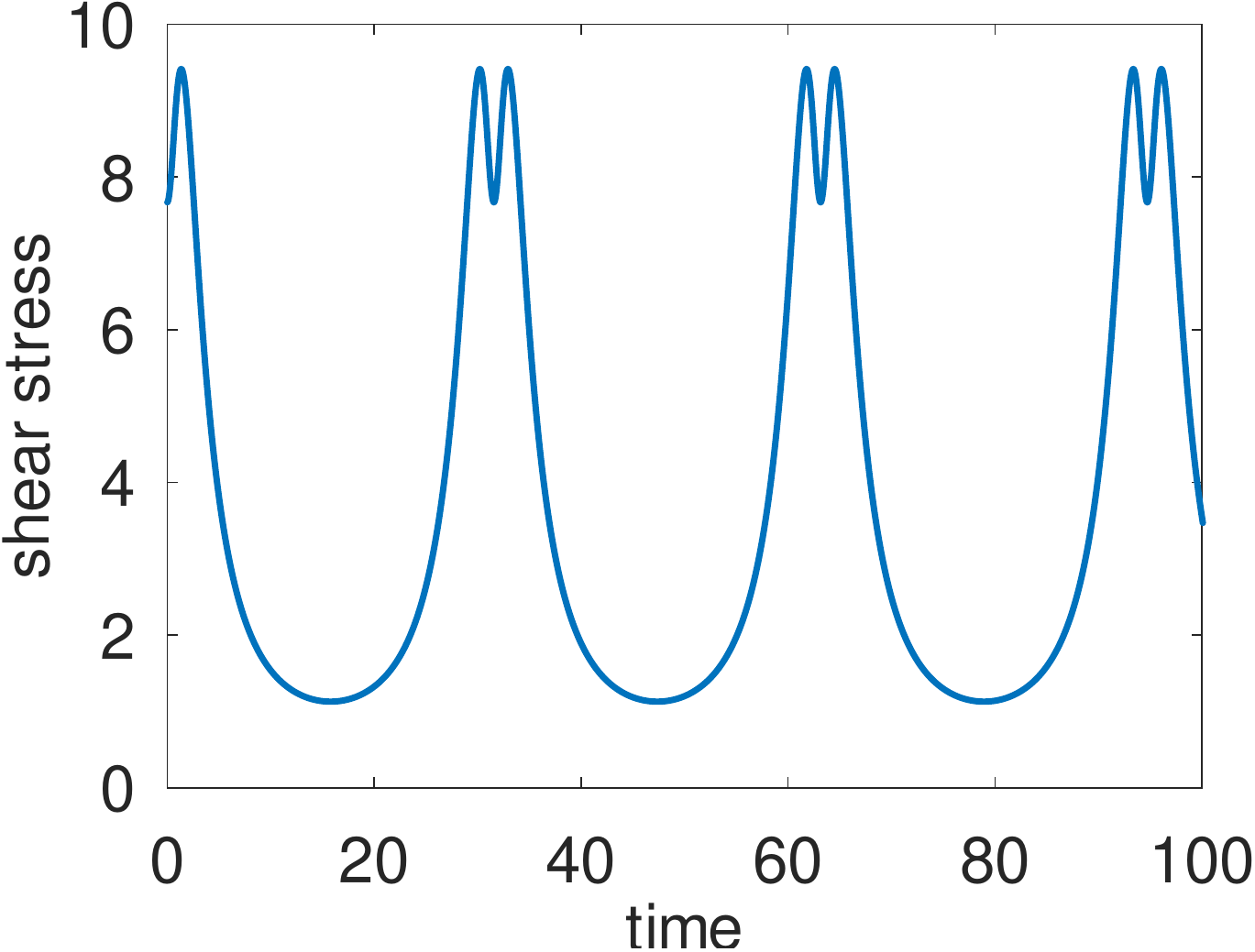}
\includegraphics[scale=0.4]{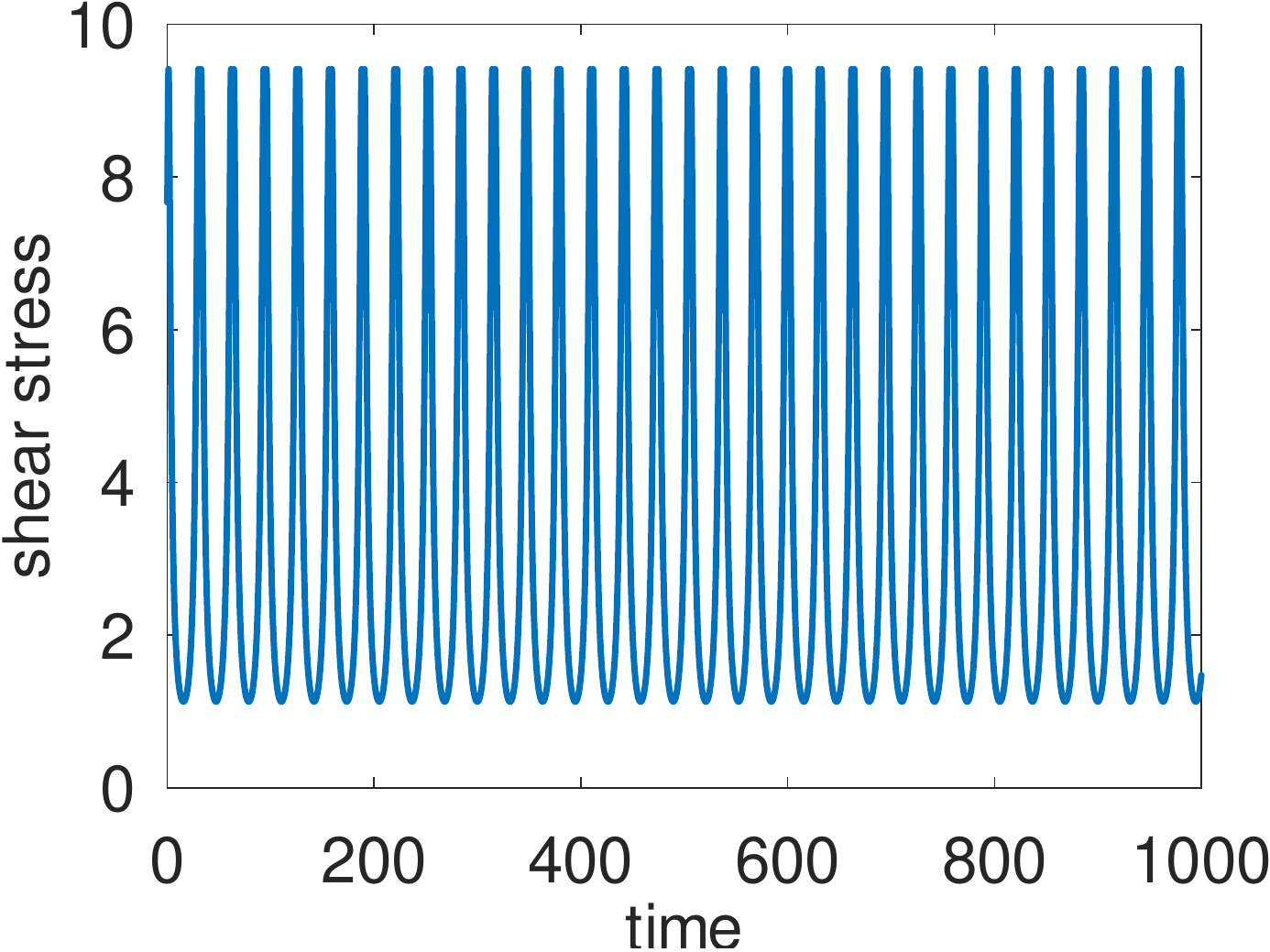}
\end{center}
\caption{Shear stress for shear flow with unperturbed fiber orientations.}
\label{fig2a}
\end{figure}

Figure~\ref{fig2a} shows the shear stress in the unperturbed case, with $\mathsf B = \mathsf I$ at $t=0$.  We give two plots, the first running for a time up to $t = 10$, and the second for a time up to $t = 100$.  The shear stress is the amount of force per unit area of plate exerted in the $x$-direction, needed to maintain the constant shear rate $V/W=1$, and is $\Sigma_1$ as defined in equation~\eqref{stokes1}.

\begin{figure}
\begin{center}
\includegraphics[scale=0.4]{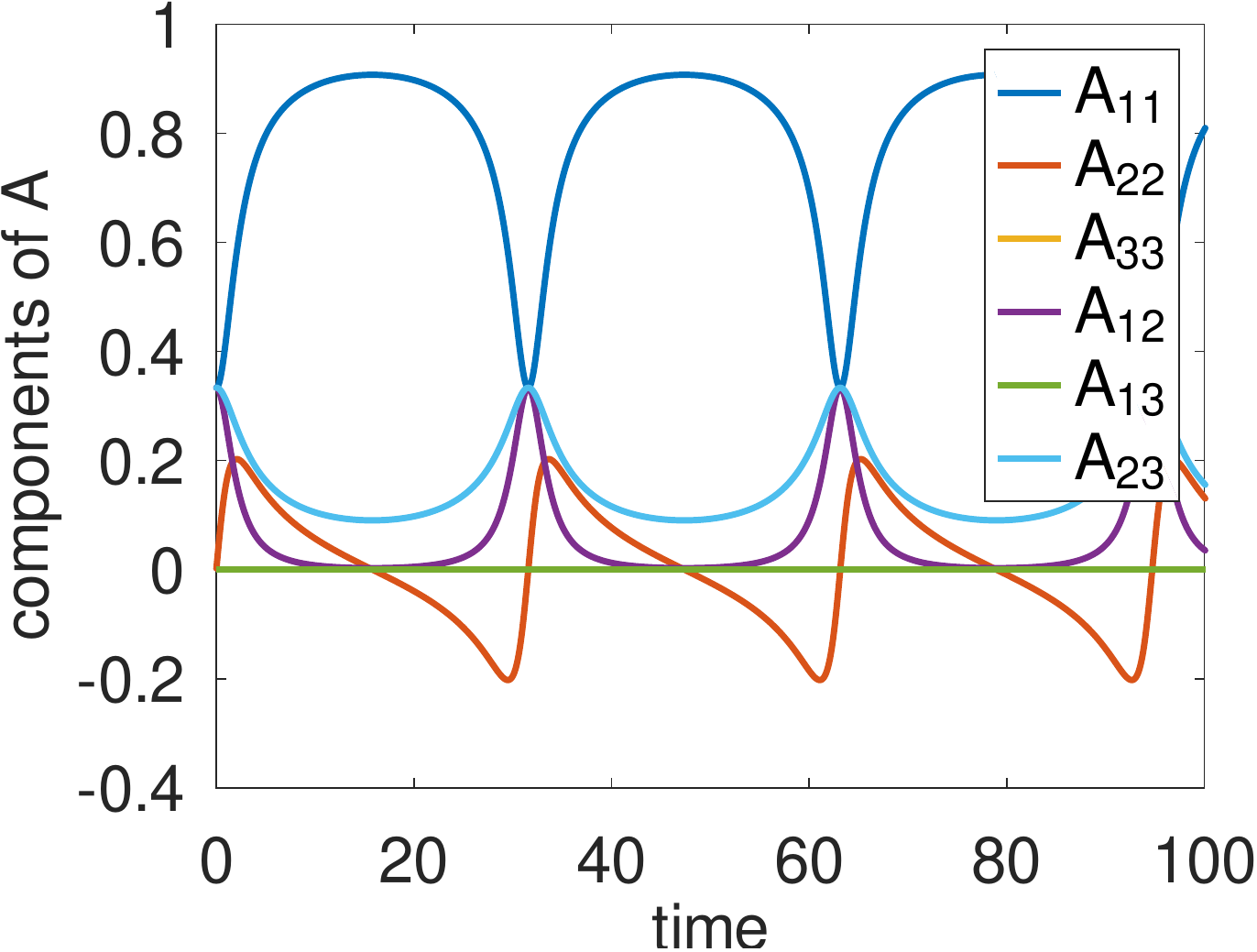}
\includegraphics[scale=0.4]{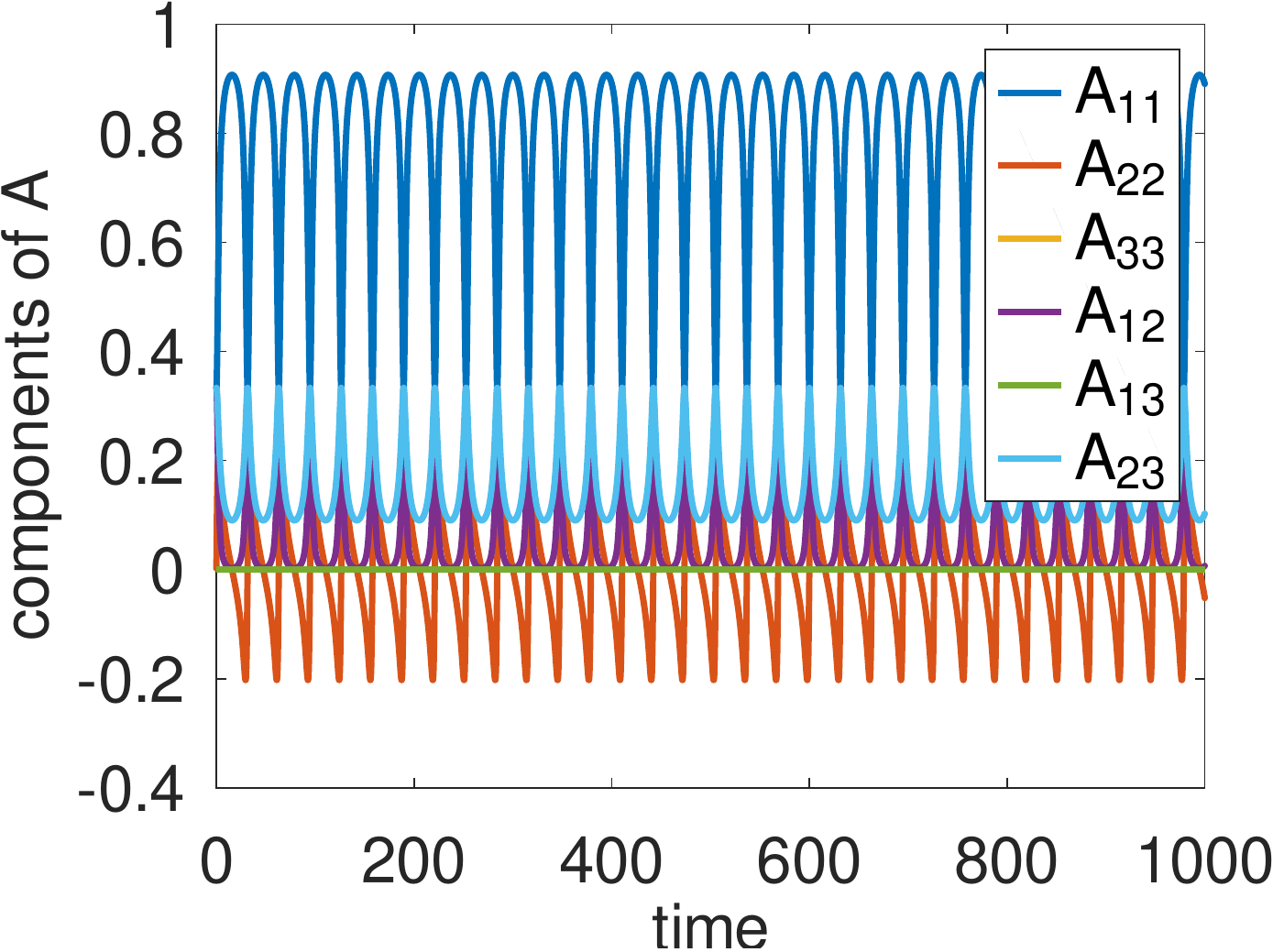}
\end{center}
\caption{Averaged over $y$ of components of $\mathsf A$ for shear flow with unperturbed fiber orientations.}
\label{fig2a-a2}
\end{figure}

Figure~\ref{fig2a-a2} shows similar plots, but for the averaged over $y$ of the components of the $\mathsf A$ matrix, which are calculated from $\mathsf B$ using the formula
\begin{equation}
\mathsf A = \tfrac12 \int_0^\infty \frac{(\mathsf B+s \mathsf I)^{-1} \, ds}{\sqrt{\text{det}(\mathsf B+s \mathsf I)}}
\end{equation}
We generally prefer reporting shear stress instead of averaged over $y$ components of $\mathsf A$, because the former is much easier to measure in experiments \cite{wang:08}.

The following plots show the shear stress, and the averaged over $y$ of the components of $\mathsf A$, when the initial value of $\mathsf B$ is given by 
\begin{equation}
\mathsf B = \kappa(\mathsf I + \epsilon \mathsf R) \qquad (t=0)
\end{equation}
where $\epsilon=0.01$ or $0.3$, and $\mathsf R$ is a symmetric matrix whose entries are independent random numbers uniformly chosen in the interval $[-1,1]$, and $\kappa$ is chosen so that $\det(\mathsf B) = 1$.  Two dimensional perturbations have the same formula, except we set $\mathsf R_{13} = \mathsf R_{23} = 0$.

\begin{figure}
\begin{center}
\includegraphics[scale=0.4]{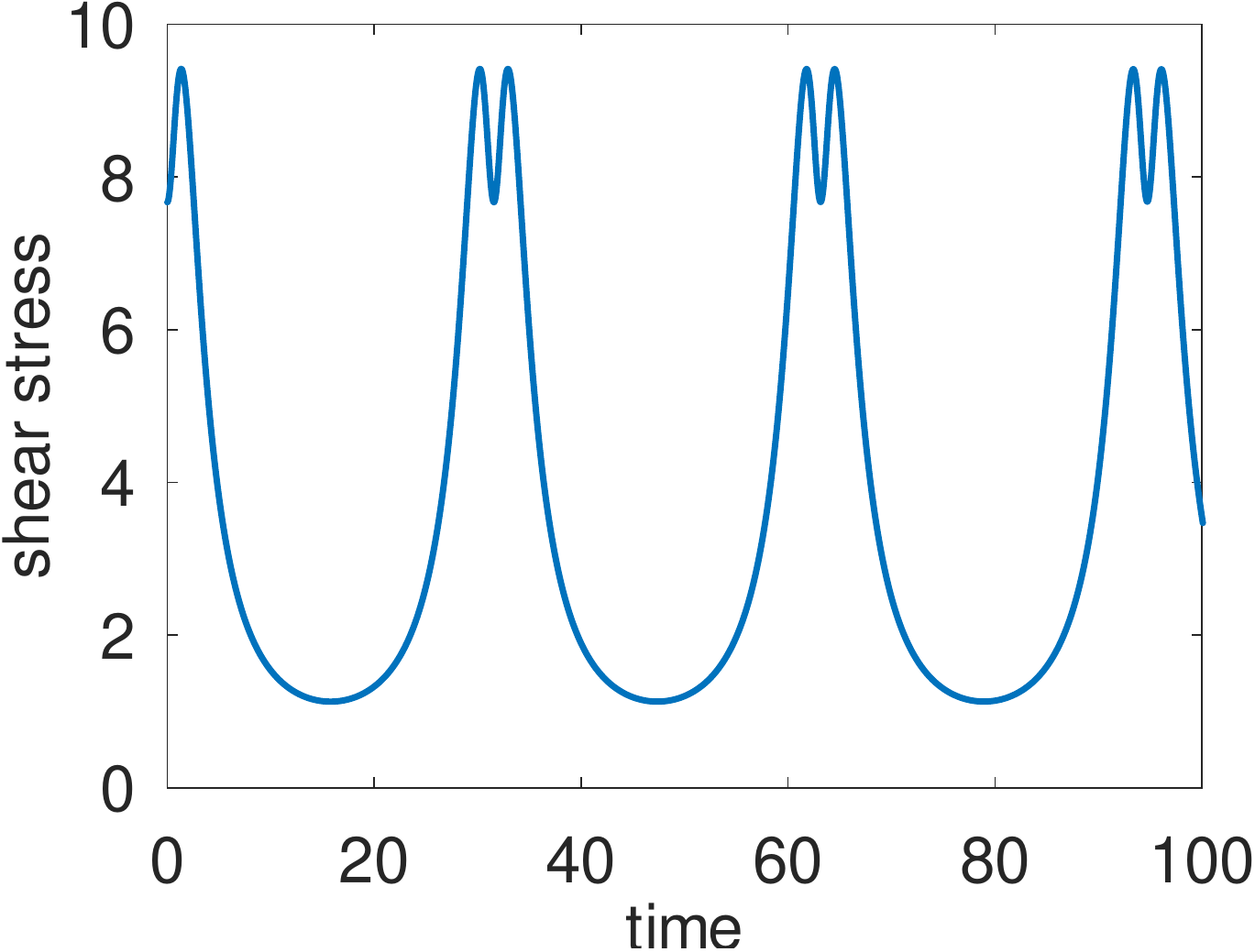}
\includegraphics[scale=0.4]{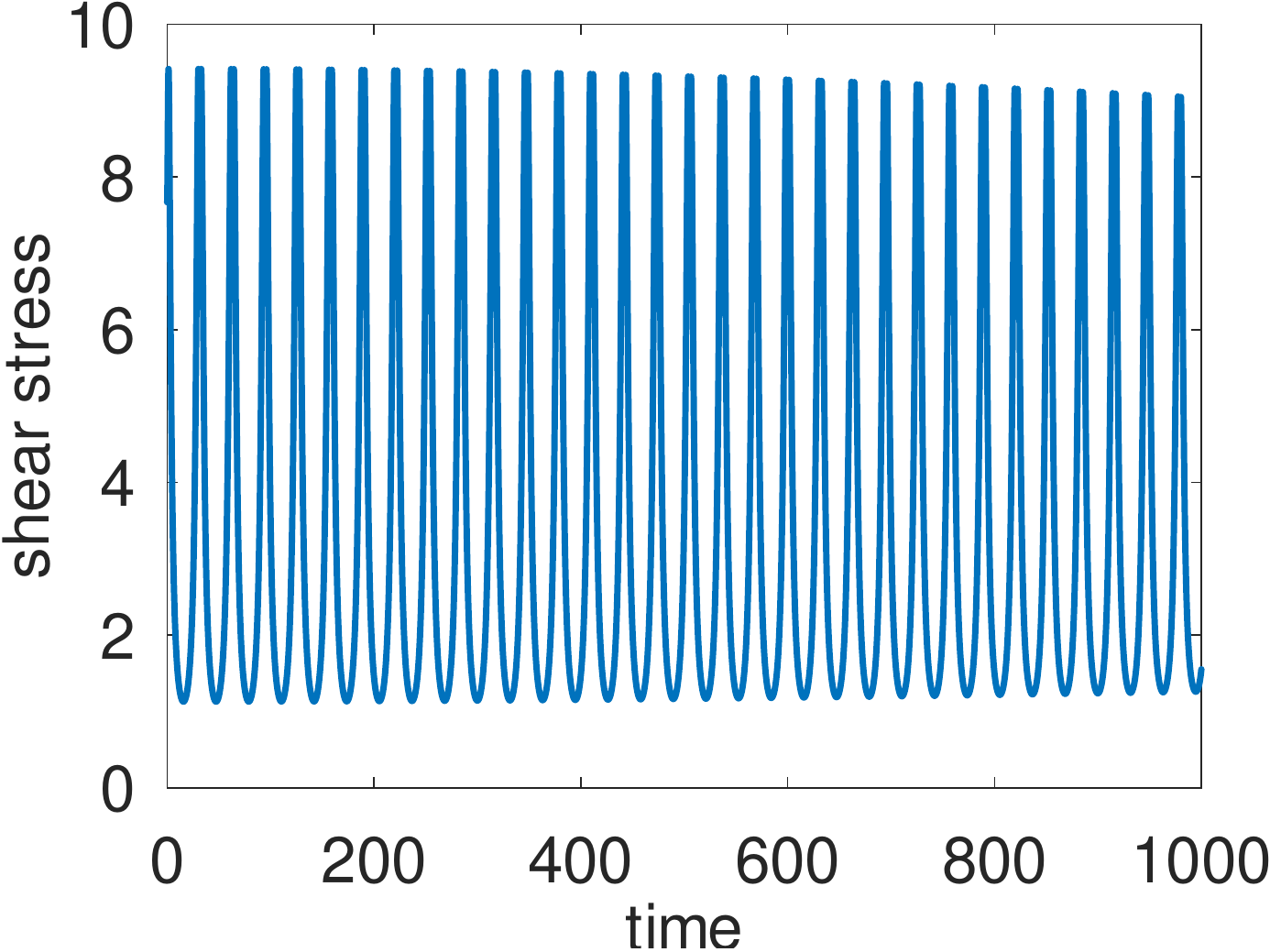}
\end{center}
\caption{Shear stress for shear flow with two dimensional fiber orientations perturbed by $\epsilon = 0.01$.}
\label{fig2b}
\end{figure}

\begin{figure}
\begin{center}
\includegraphics[scale=0.4]{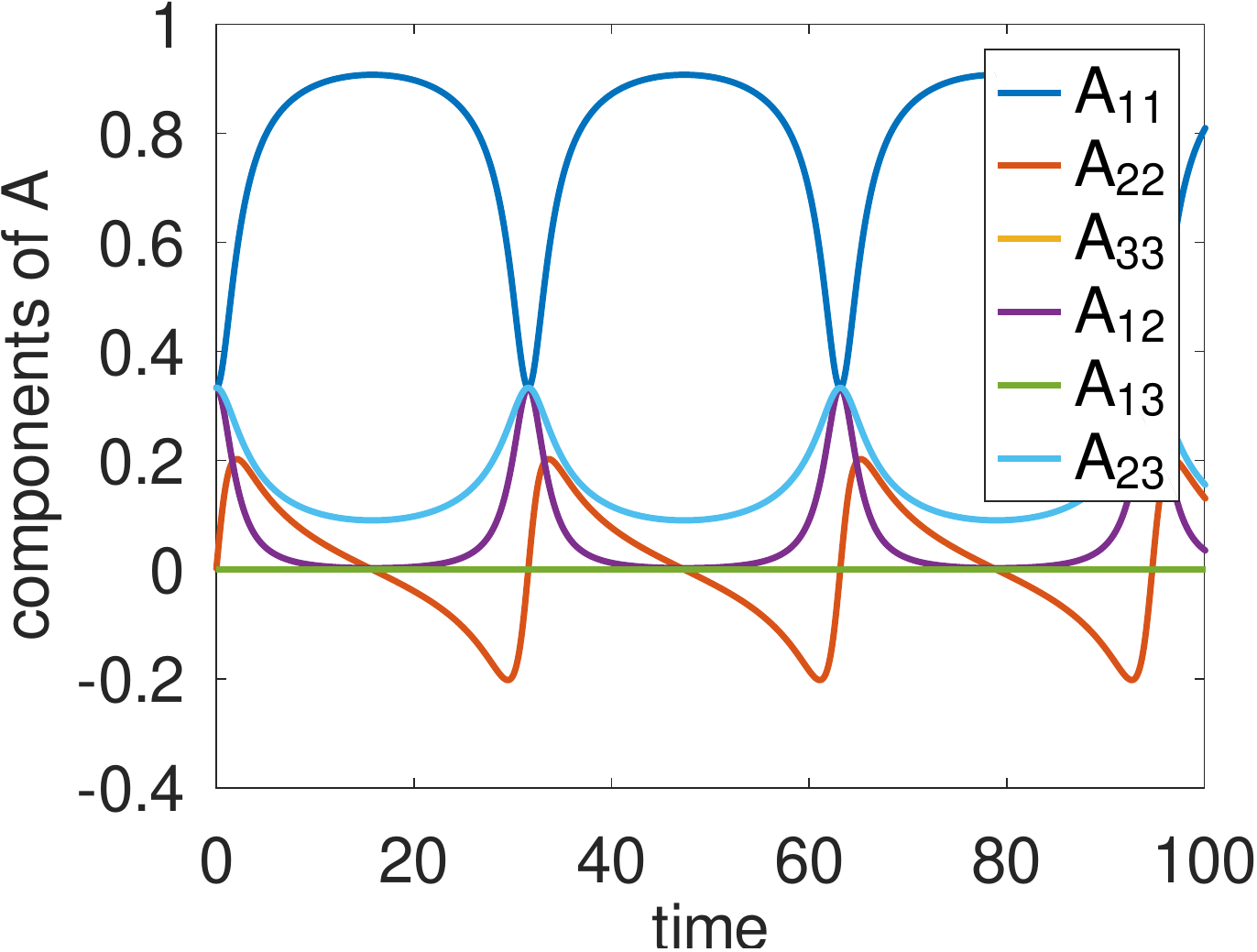}
\includegraphics[scale=0.4]{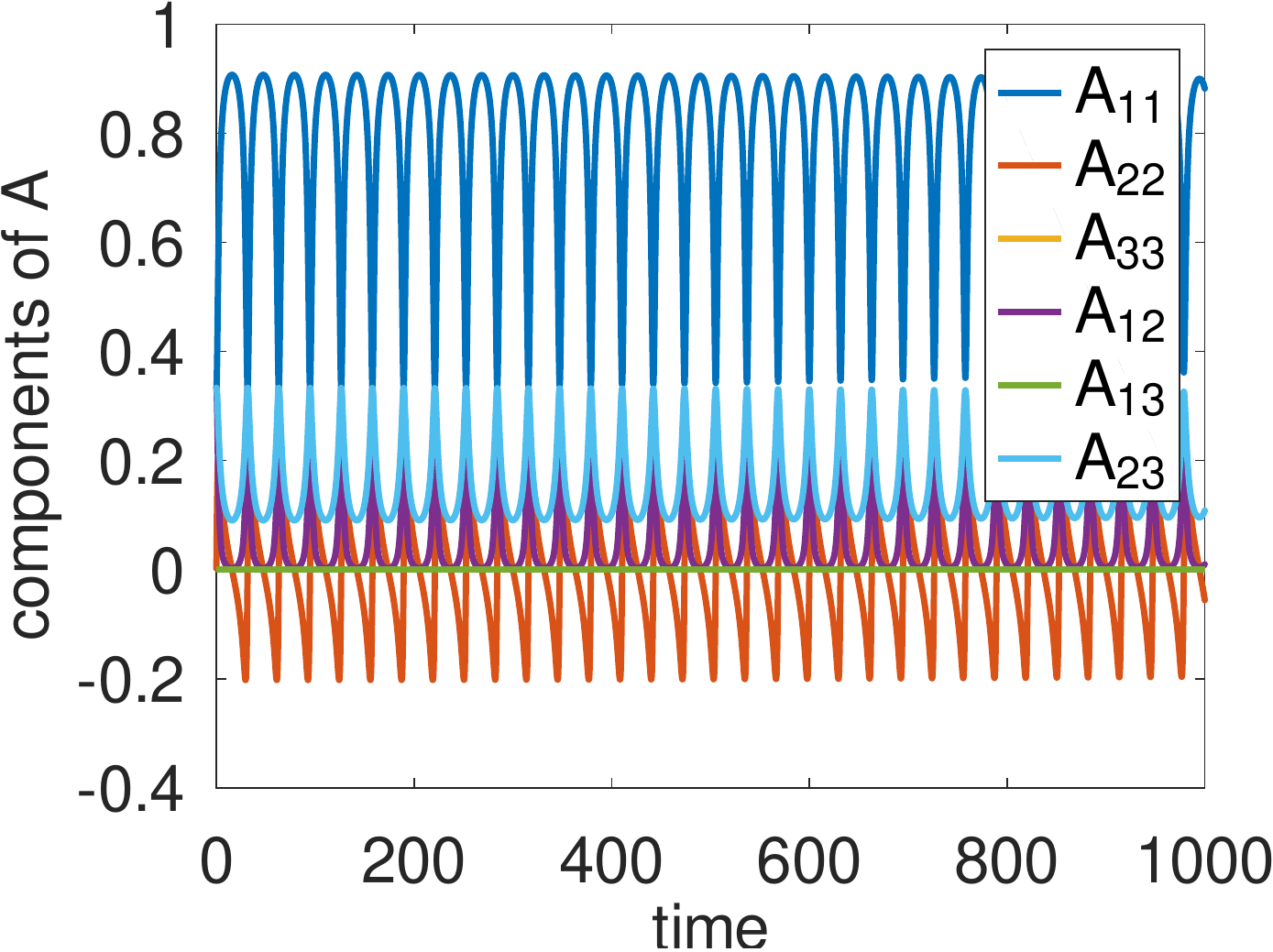}
\end{center}
\caption{Averaged over $y$ of components of $\mathsf A$ for shear flow with two dimensional fiber orientations perturbed by $\epsilon = 0.01$.}
\label{fig2b-a2}
\end{figure}

\begin{figure}
\begin{center}
\includegraphics[scale=0.4]{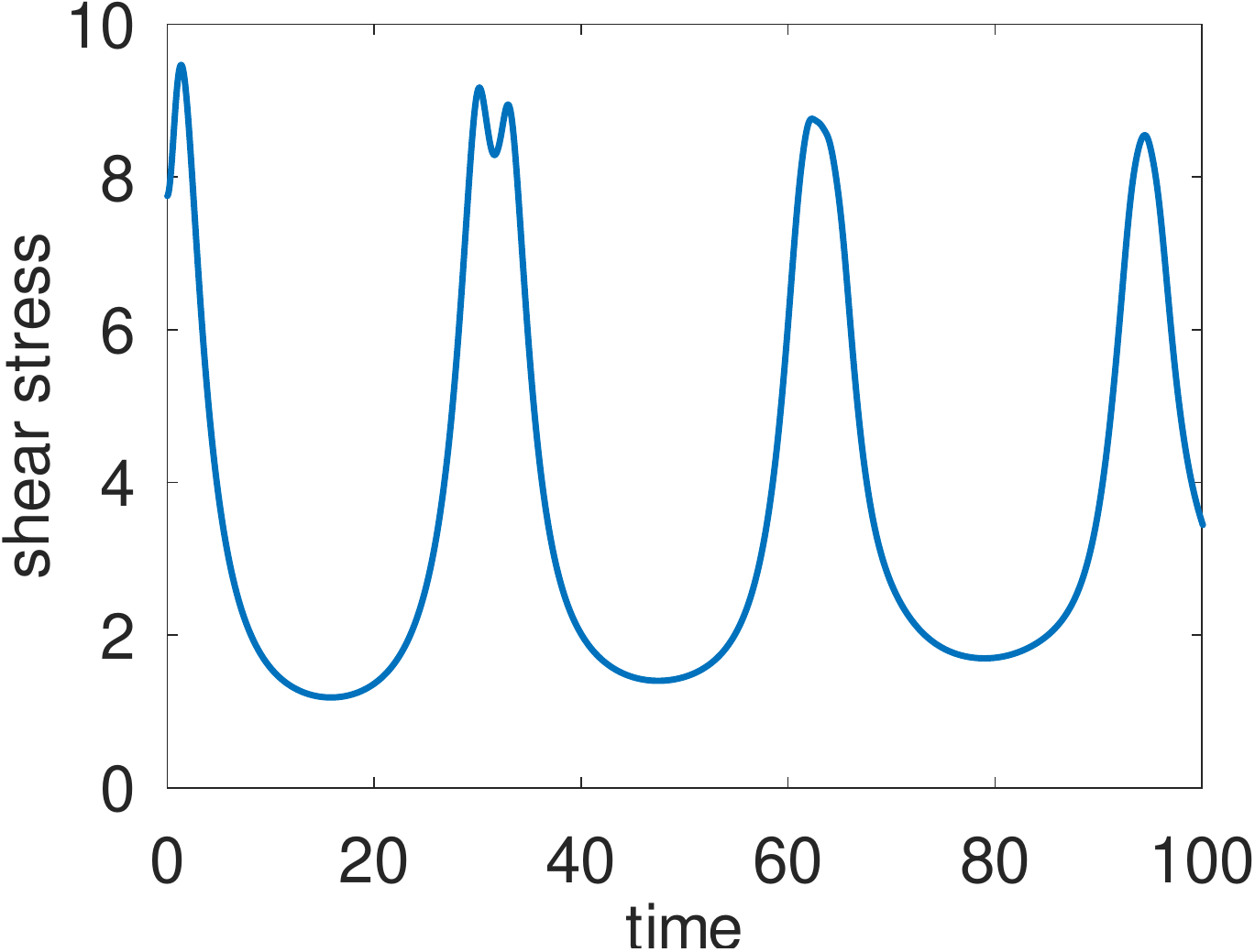}
\includegraphics[scale=0.4]{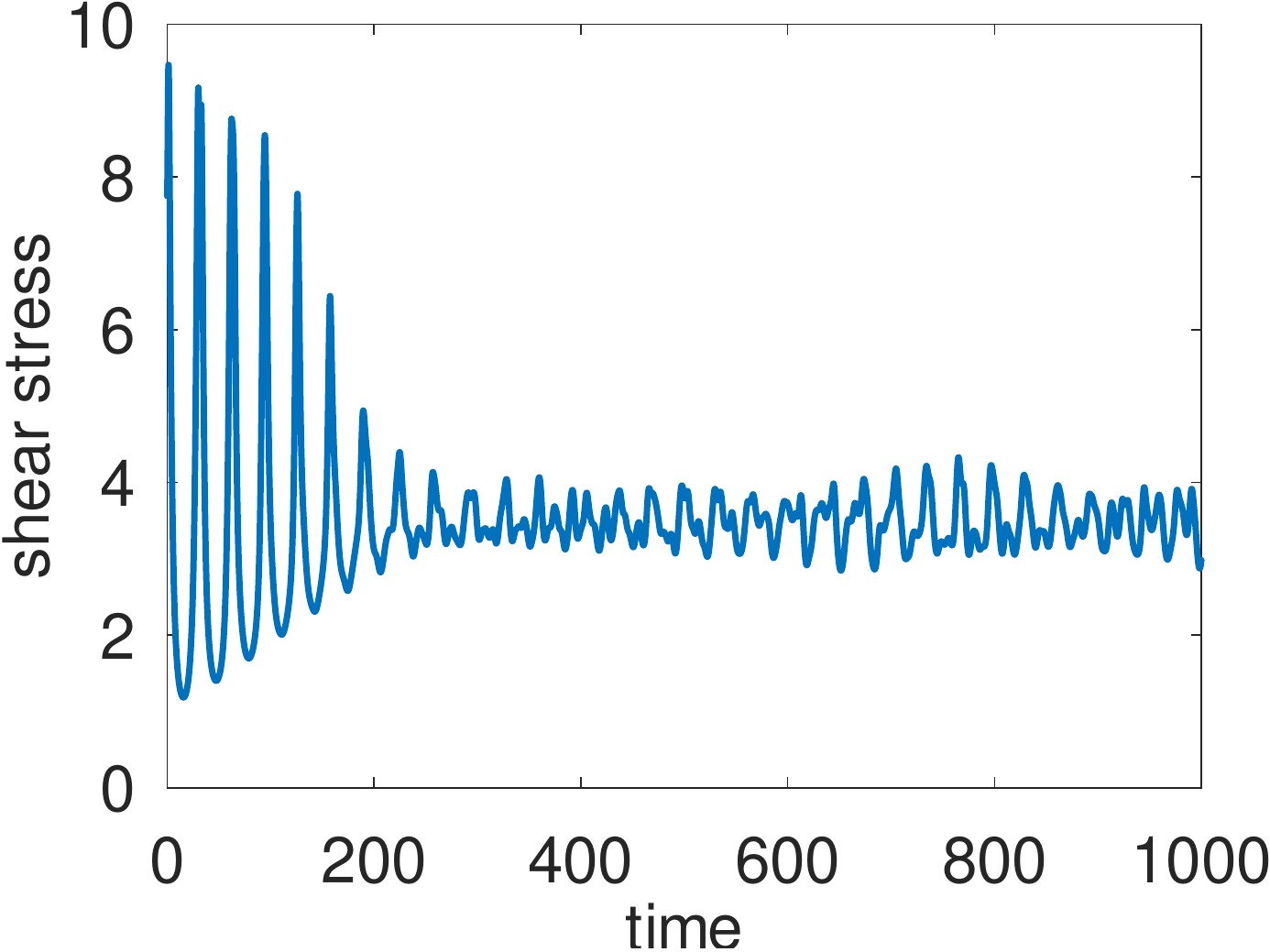}
\end{center}
\caption{Shear stress for shear flow with two dimensional fiber orientations perturbed by $\epsilon = 0.3$.}
\label{fig2c}
\end{figure}

\begin{figure}
\begin{center}
\includegraphics[scale=0.4]{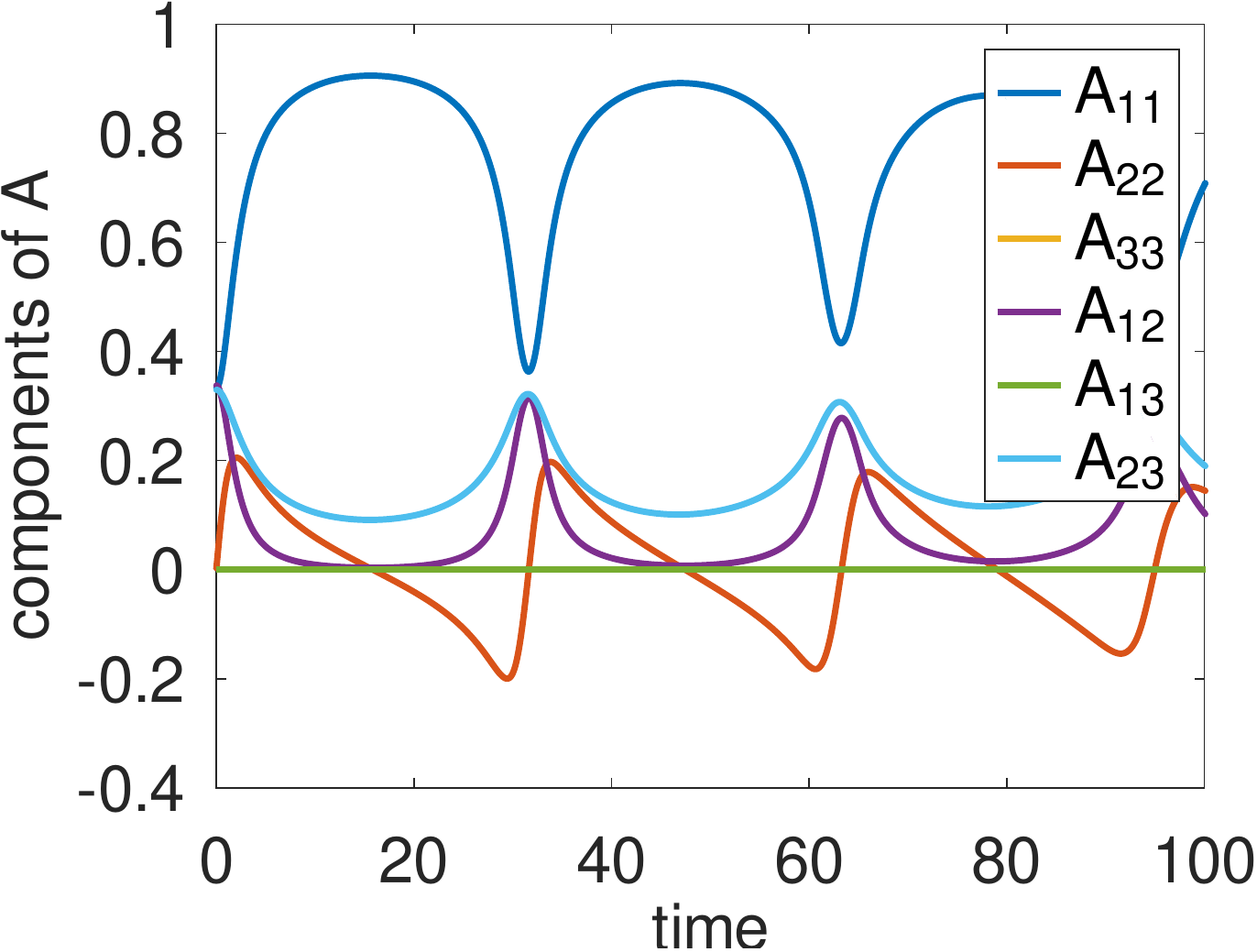}
\includegraphics[scale=0.4]{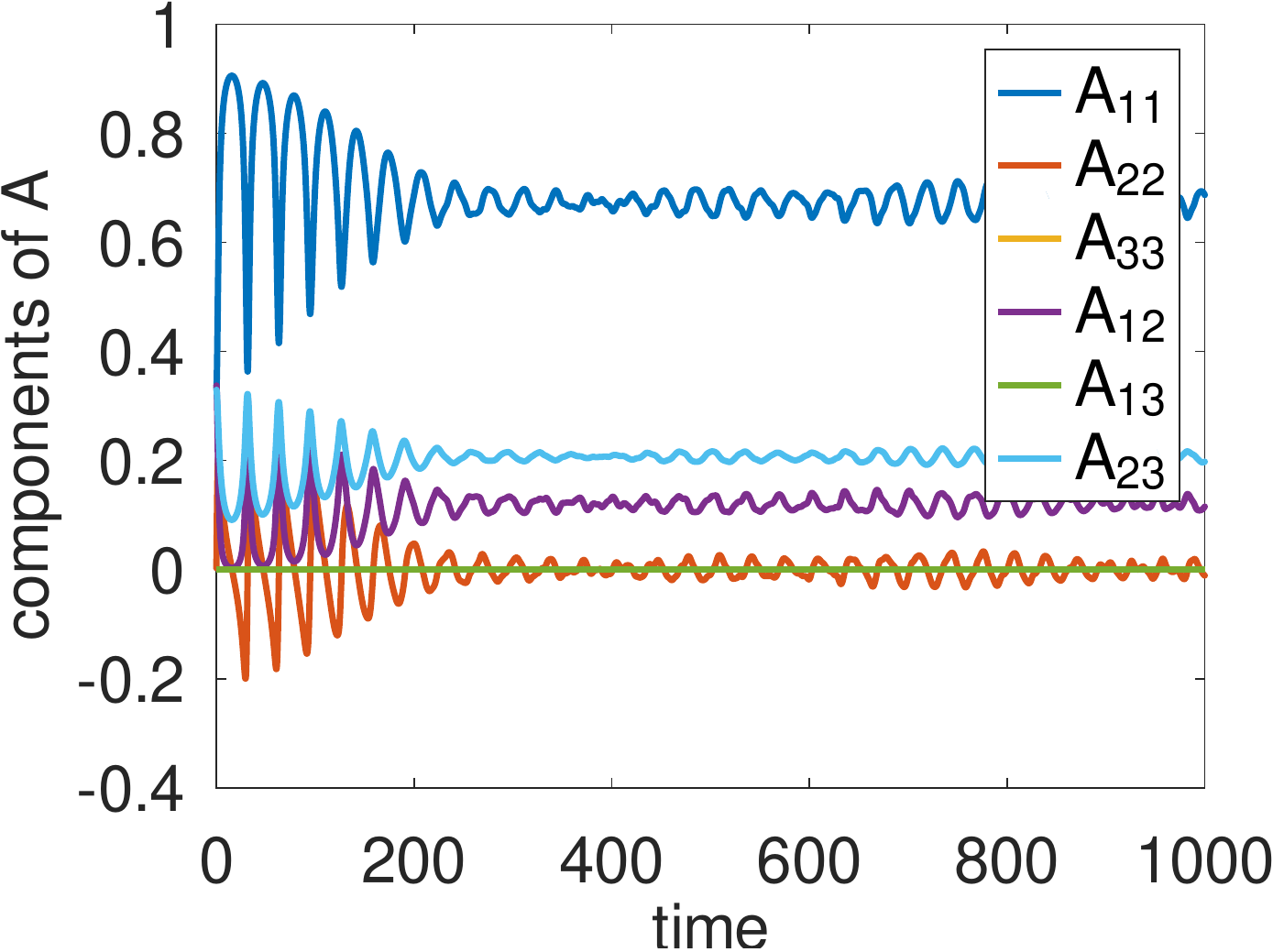}
\end{center}
\caption{Averaged over $y$ of components of $\mathsf A$ for shear flow with two dimensional fiber orientations perturbed by $\epsilon = 0.3$.}
\label{fig2c-a2}
\end{figure}

Figures~\ref{fig2b} and~\ref{fig2b-a2} show the shear stress and averaged over $y$ of the components of $\mathsf A$ for a two dimensional perturbation with $\epsilon = 0.01$,
Figures~\ref{fig2c} and~\ref{fig2c-a2} show the shear stress and averaged over $y$ of the components of $\mathsf A$ for a two dimensional perturbation with $\epsilon = 0.3$, 
Figures~\ref{fig3c} and~\ref{fig3c-a2} show the shear stress and averaged over $y$ of the components of $\mathsf A$ for an unrestricted perturbation with $\epsilon = 0.3$, 
and
Figures~\ref{fig2c} and~\ref{fig2c-a2} show the shear stress and averaged over $y$ of the components of $\mathsf A$ for an unrestricted perturbation with $\epsilon = 0.3$.

Figures~\ref{fig2b-verylong} and~\ref{fig3b-verylong} show the two dimensional and restricted perturbations with $\epsilon = 0.01$, run up to $t=10000$.

These do not achieve an actual steady state, but after some time, the oscillations seem to cease to decrease.  for the cases $\epsilon = 0.01$ and $\epsilon = 0.3$, for unrestricted perturbations with $\lambda = 0.98$, we computed the mean and standard deviations over the second half of their respective time intervals (that is $[5000, 10000]$ and $[500, 1000]$ respectively), of both the shear stress and the averaged over $y$ of the components of $\mathsf A$,   These are shown in Table~\ref{steady-state}.  These values match quite closely.

We also provide plots that give a sense of how long it takes for the values to settle down.  We performed a moving standard deviation, with a window the size of $t_p$ based upon equation~\eqref{t_p}, upon the shear stress (see Figure~\ref{moving sd}, which shows the moving standard deviation for an unrestricted perturbation with $\lambda = 0.98$ and $\epsilon = 0.45$).  We computed how long it took for this moving standard deviation to become less than $1$, and these times to settle are plotted in Figures~\ref{time-to-settle-eps} and~\ref{time-to-settle-lambda}, with both the two dimensional perturbations and the unrestricted perturbations shown.  Note that the times smaller than about $100$ may be partially effected the value of $t_p$, which is comparable, and thus these times should be considered suspect.

Finally, note that these results only verify that the solutions are unstable.  Furthermore, since the flows are only allowed to be shear flows, these results shouldn't be seen as indicative of what may happen in experiments, as a more general perturbation will break up the shear flow.

We conjecture that if a full three dimensional simulation is run, which allows for any fluid motion, not just shear flow, then the solution will converge to a steady state much more rapidly in the case of unrestricted perturbations.

\begin{figure}
\begin{center}
\includegraphics[scale=0.4]{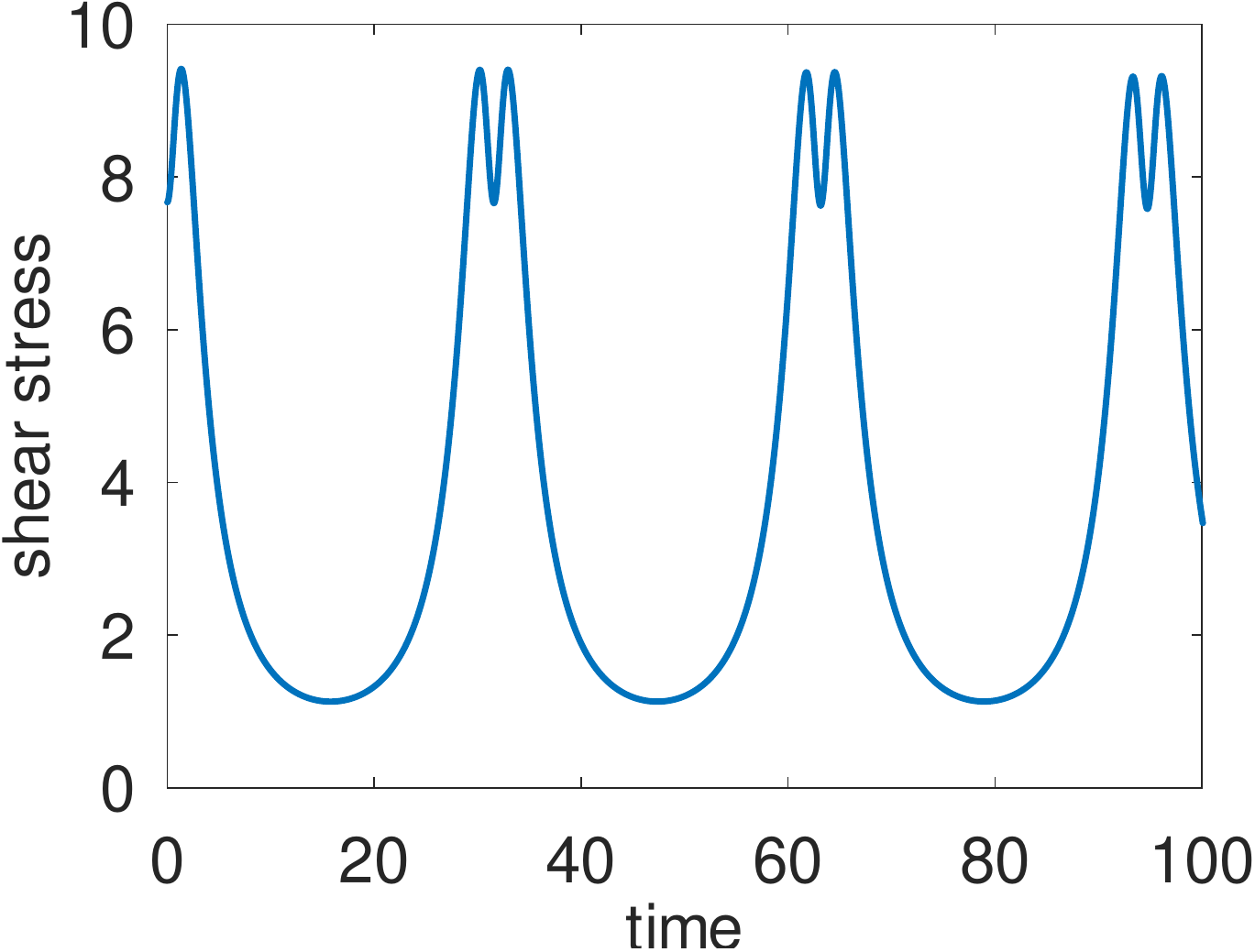}
\includegraphics[scale=0.4]{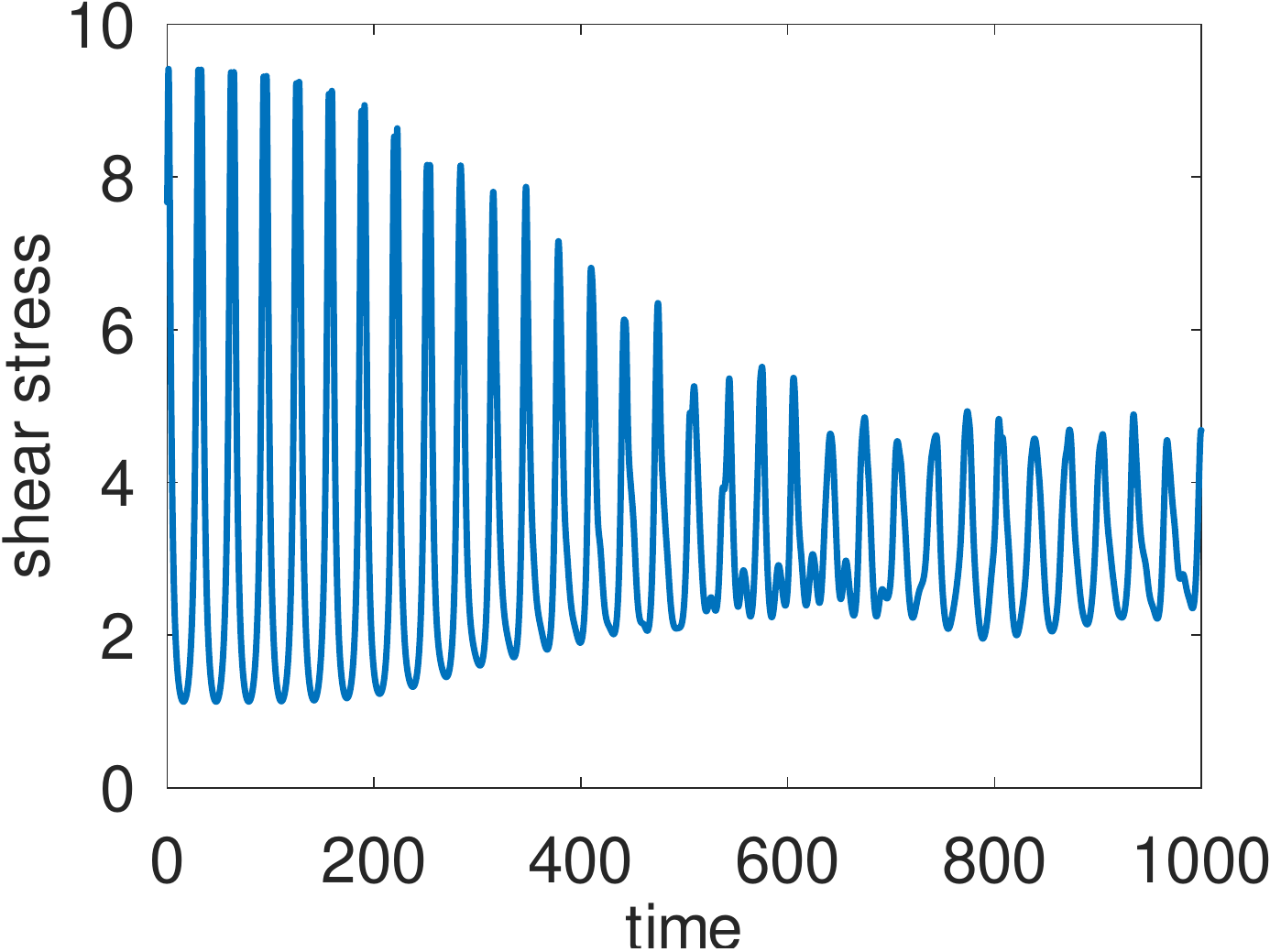}
\end{center}
\caption{Shear stress for shear flow with unrestricted fiber orientations perturbed by $\epsilon = 0.01$.}
\label{fig3b}
\end{figure}

\begin{figure}
\begin{center}
\includegraphics[scale=0.4]{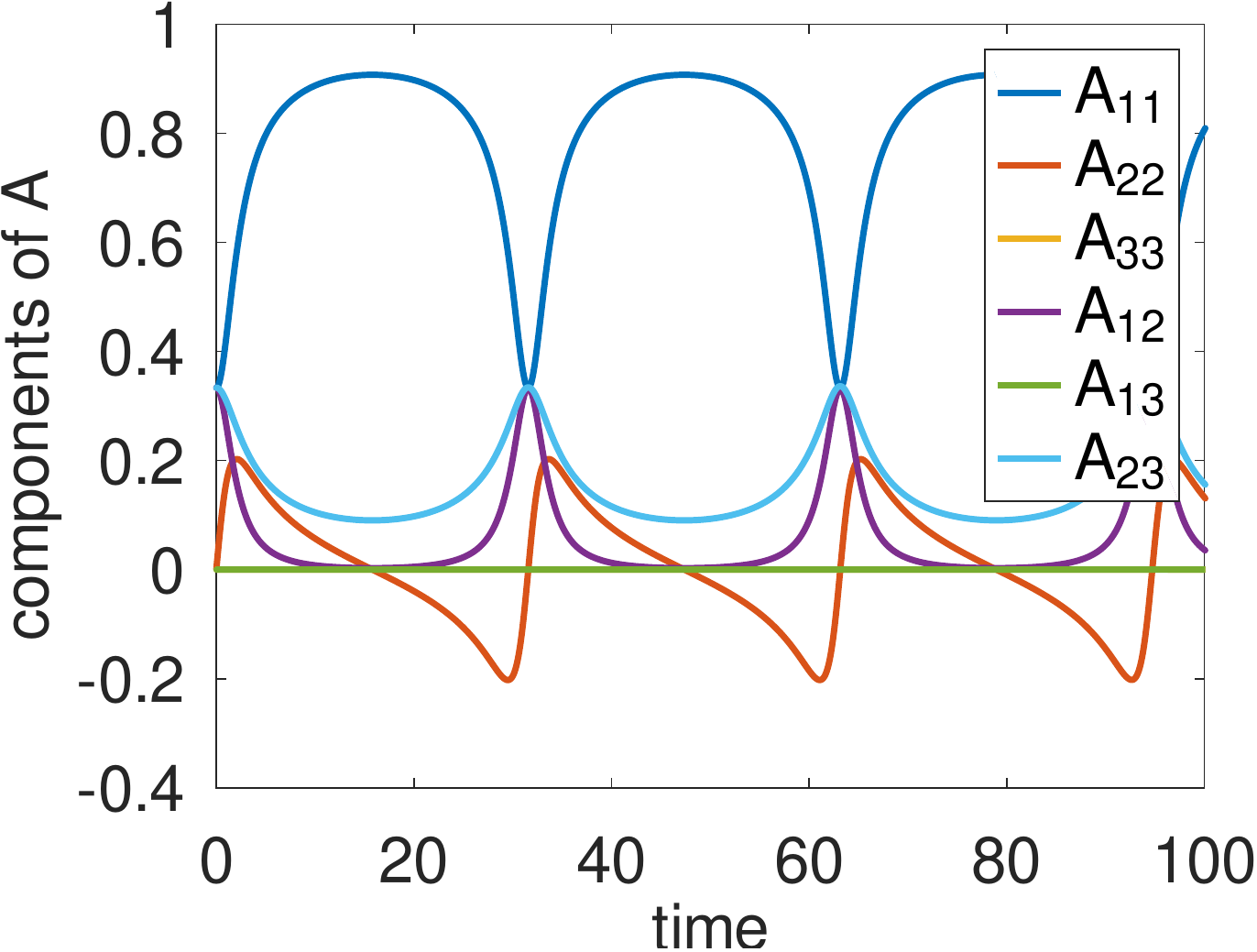}
\includegraphics[scale=0.4]{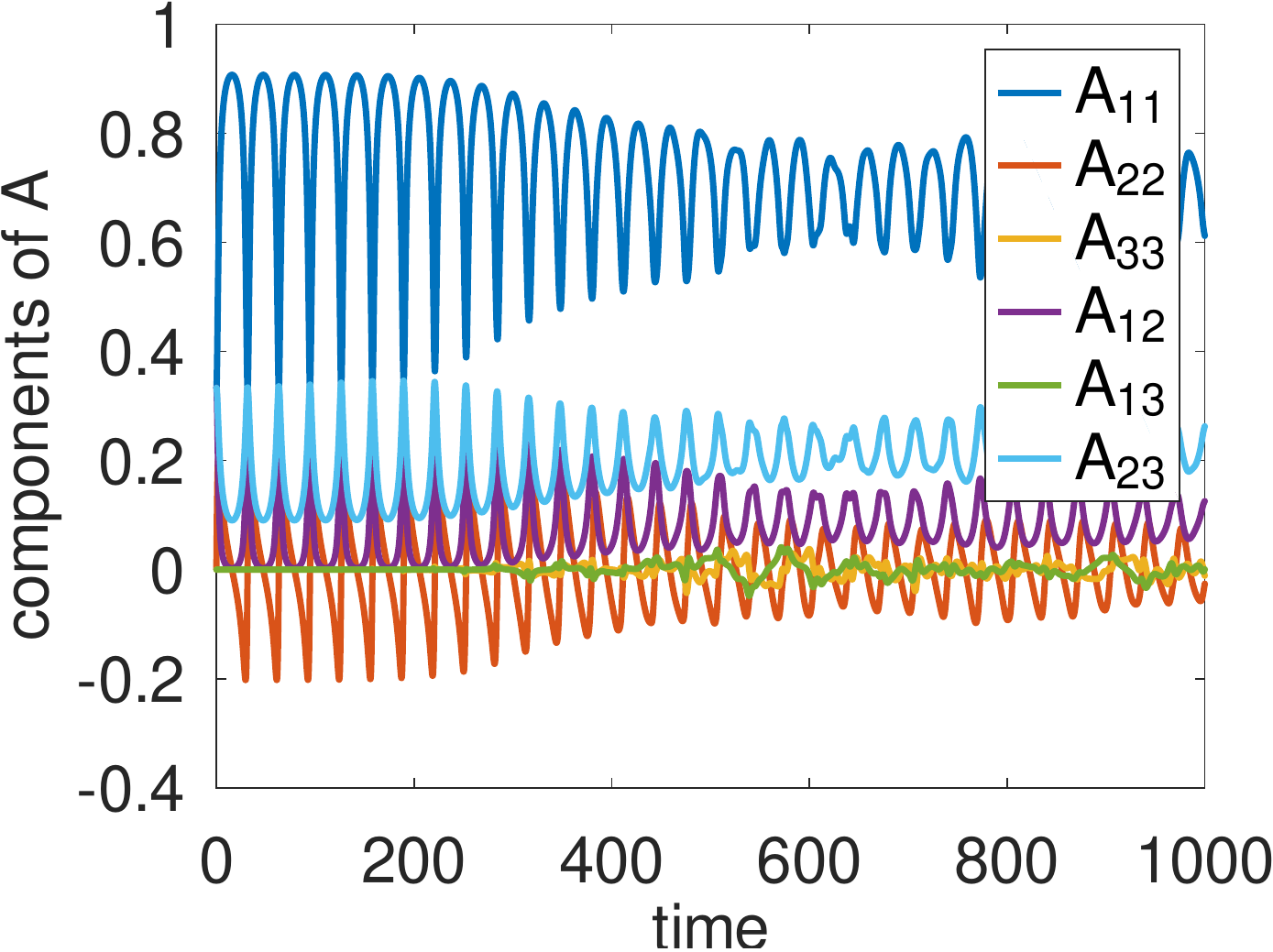}
\end{center}
\caption{Averaged over $y$ of components of $\mathsf A$ for shear flow with unrestricted fiber orientations perturbed by $\epsilon = 0.01$.}
\label{fig3b-a2}
\end{figure}

\begin{figure}
\begin{center}
\includegraphics[scale=0.4]{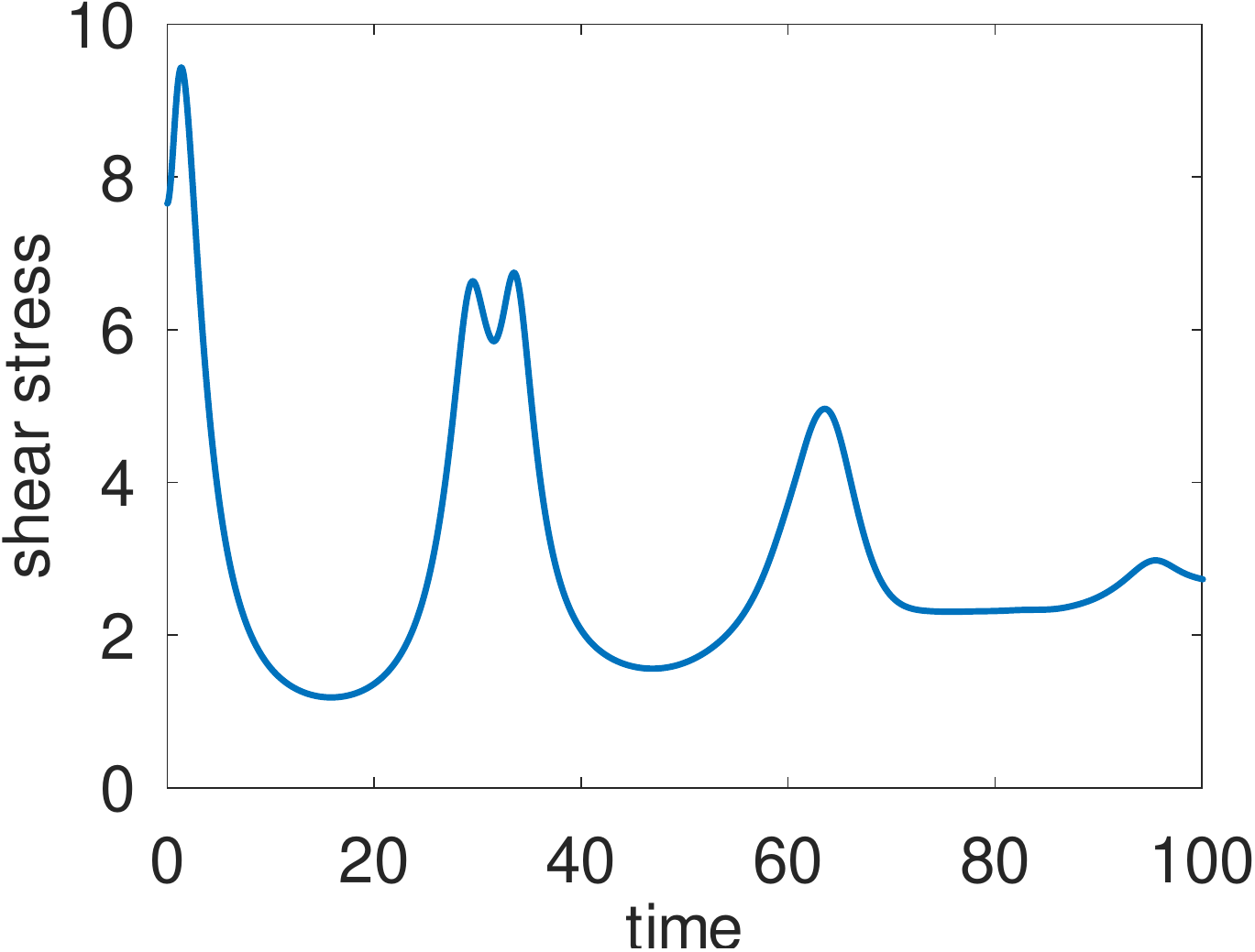}
\includegraphics[scale=0.4]{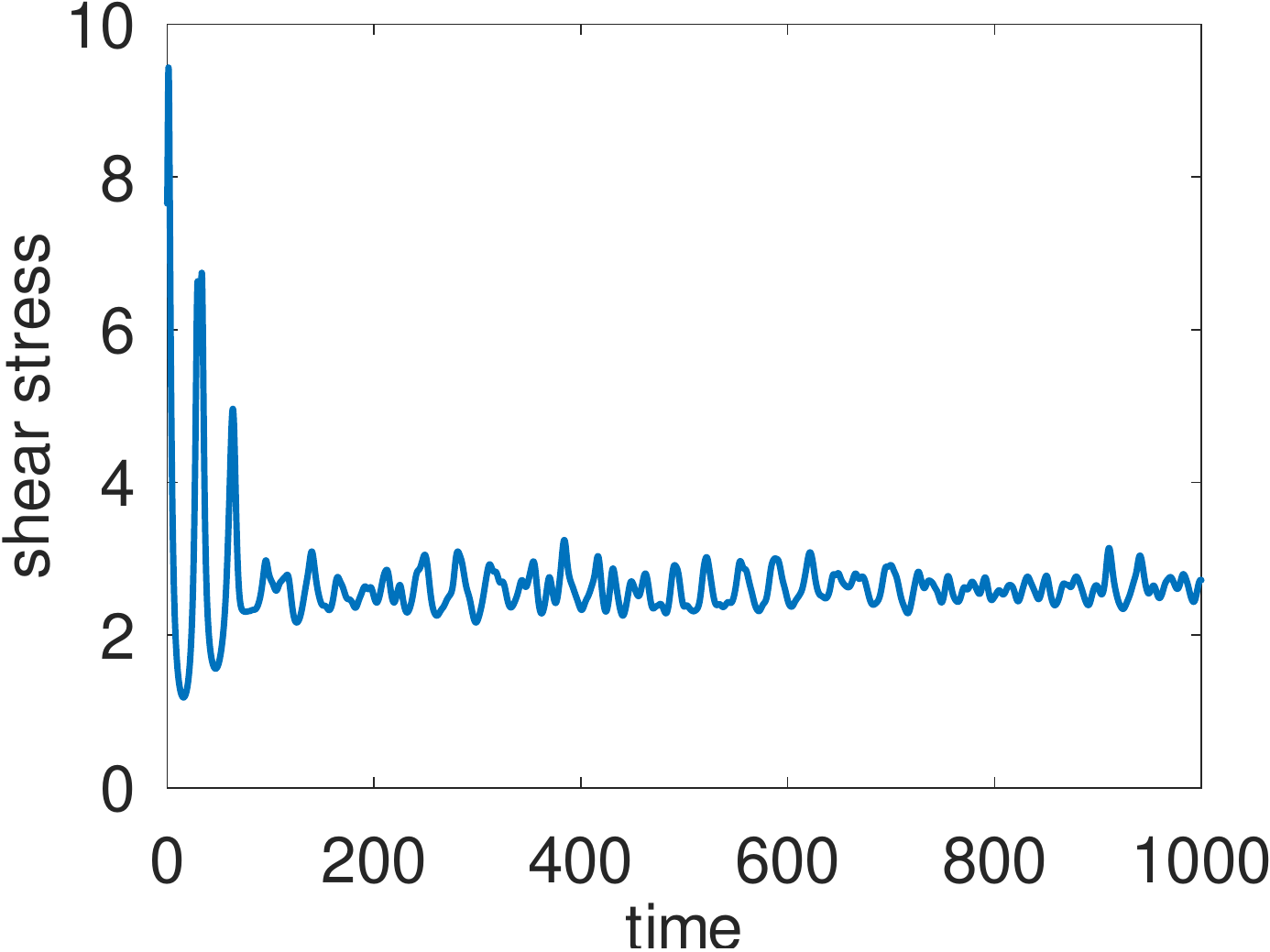}
\end{center}
\caption{Shear stress for shear flow with unrestricted fiber orientations perturbed by $\epsilon = 0.3$.}
\label{fig3c}
\end{figure}

\begin{figure}
\begin{center}
\includegraphics[scale=0.4]{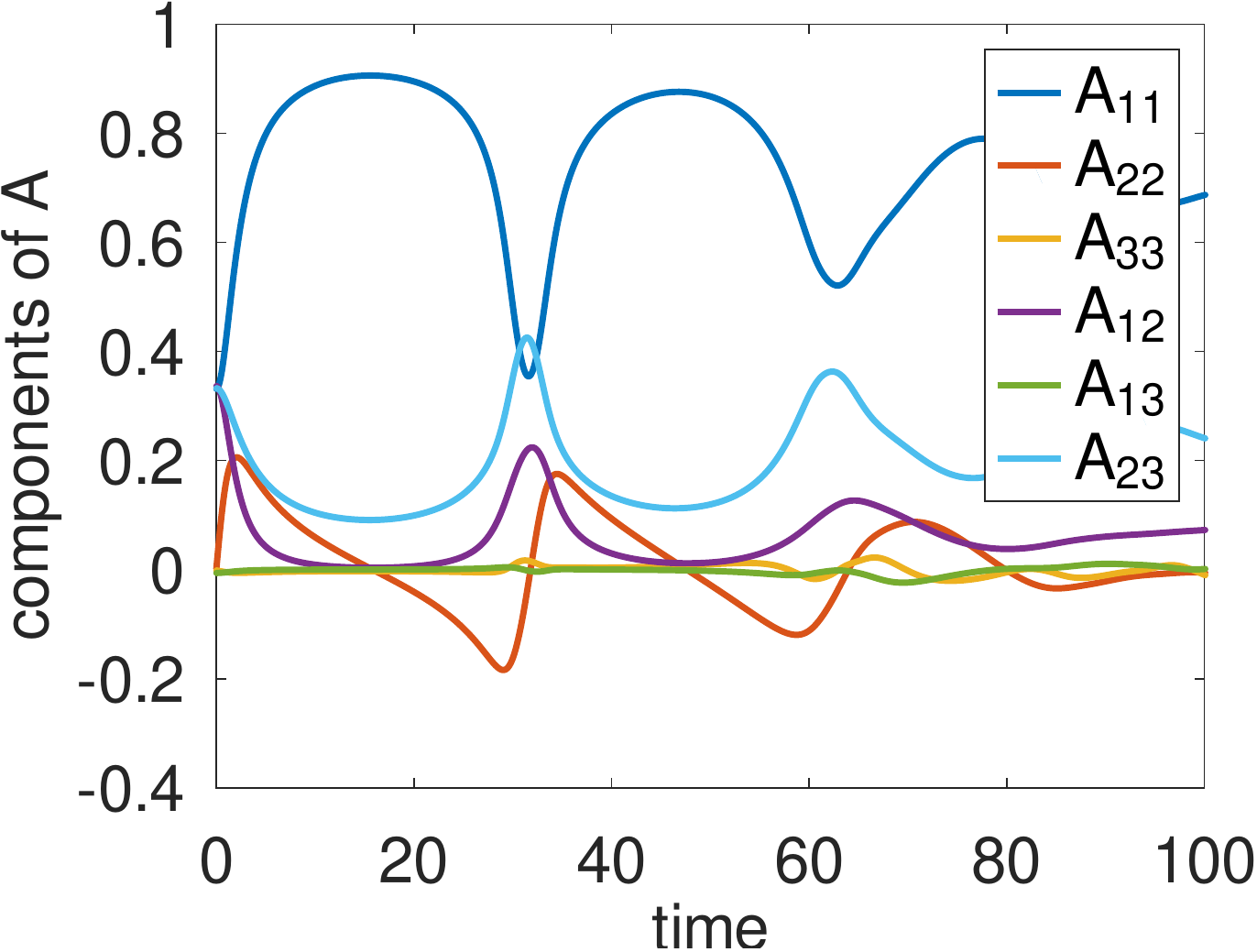}
\includegraphics[scale=0.4]{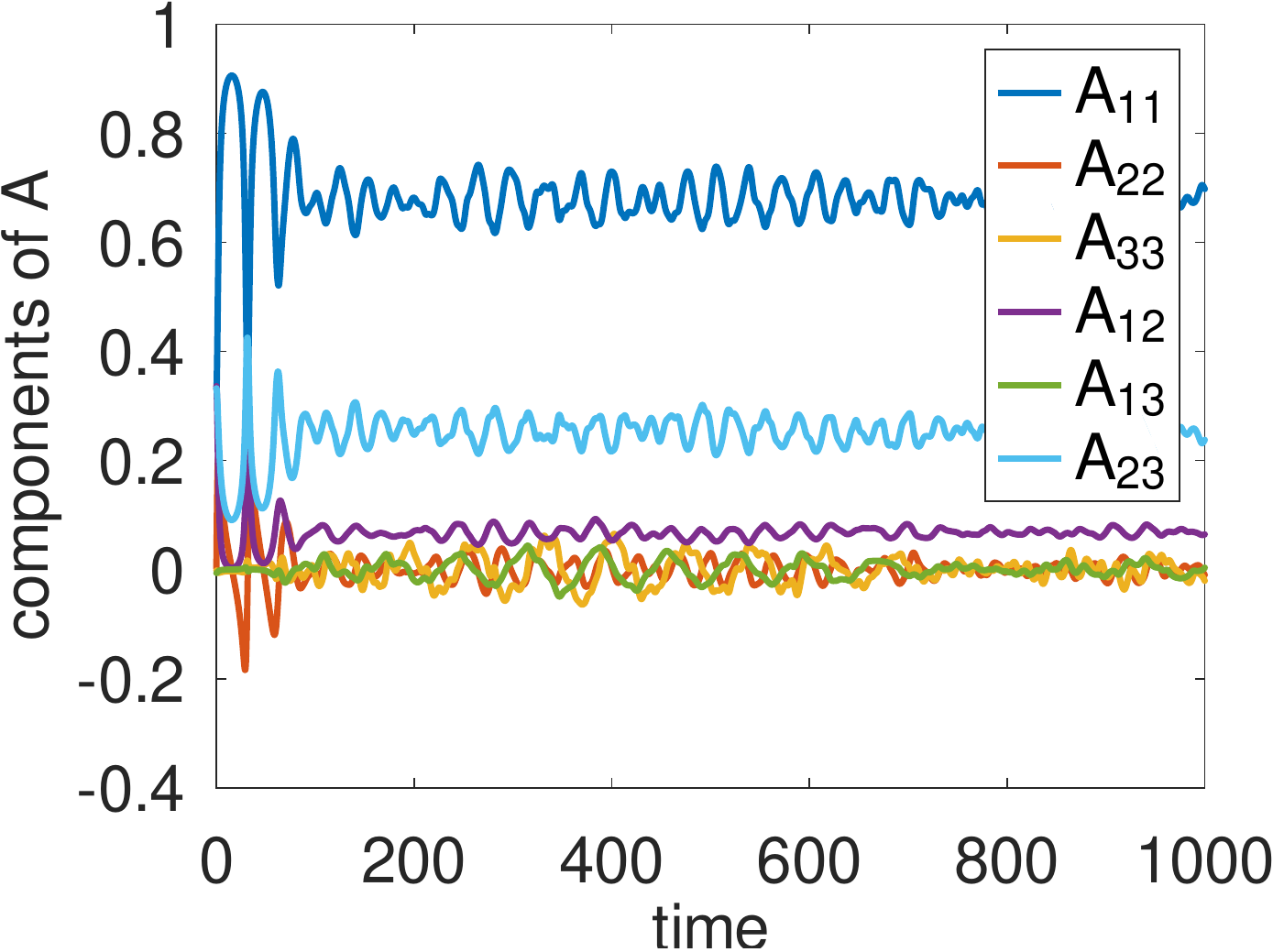}
\end{center}
\caption{Averaged over $y$ of components of $\mathsf A$ for shear flow with fiber orientations perturbed by $\epsilon = 0.3$.}
\label{fig3c-a2}
\end{figure}

\begin{figure}
\begin{center}
\includegraphics[scale=0.4]{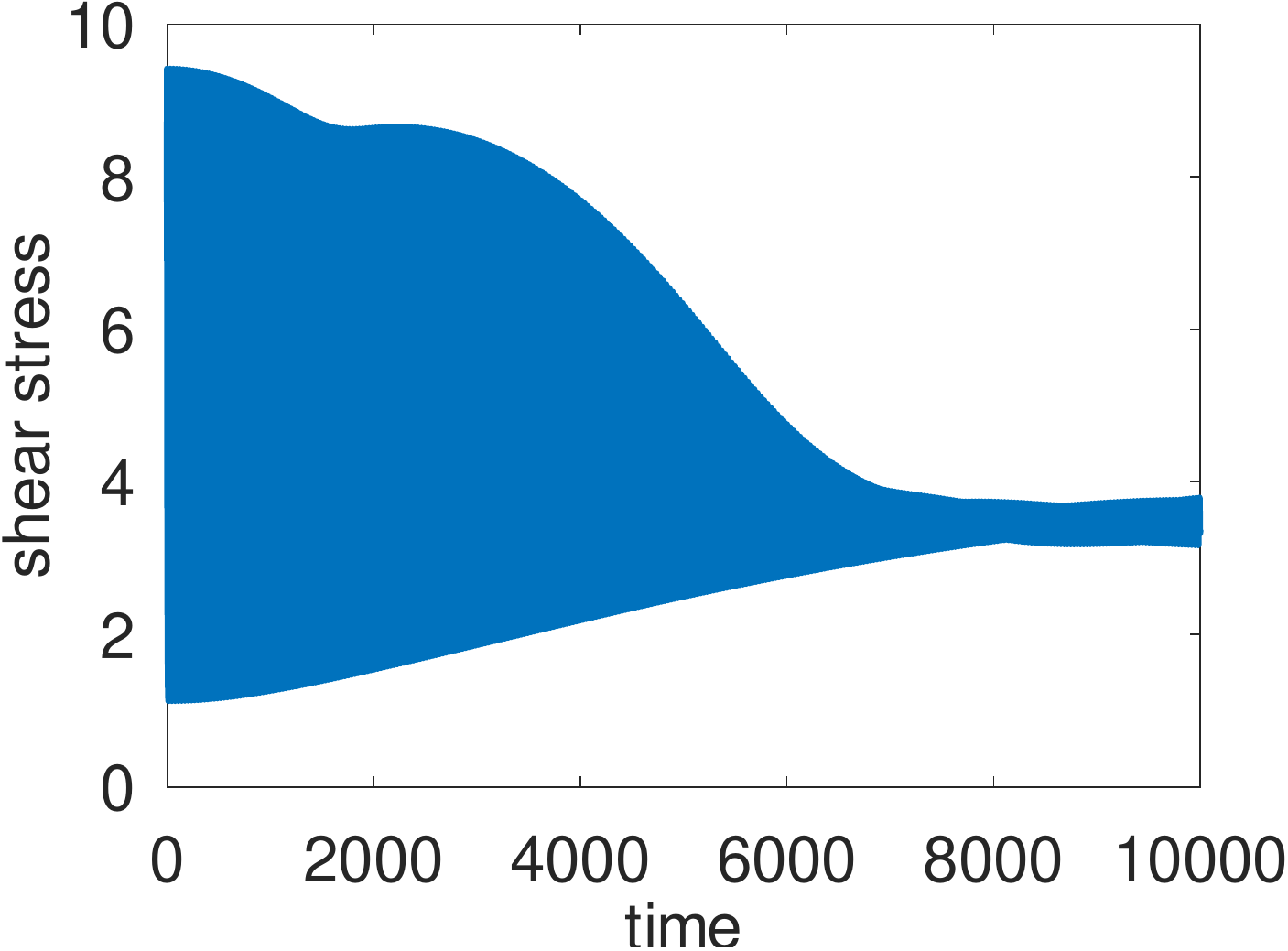}
\includegraphics[scale=0.4]{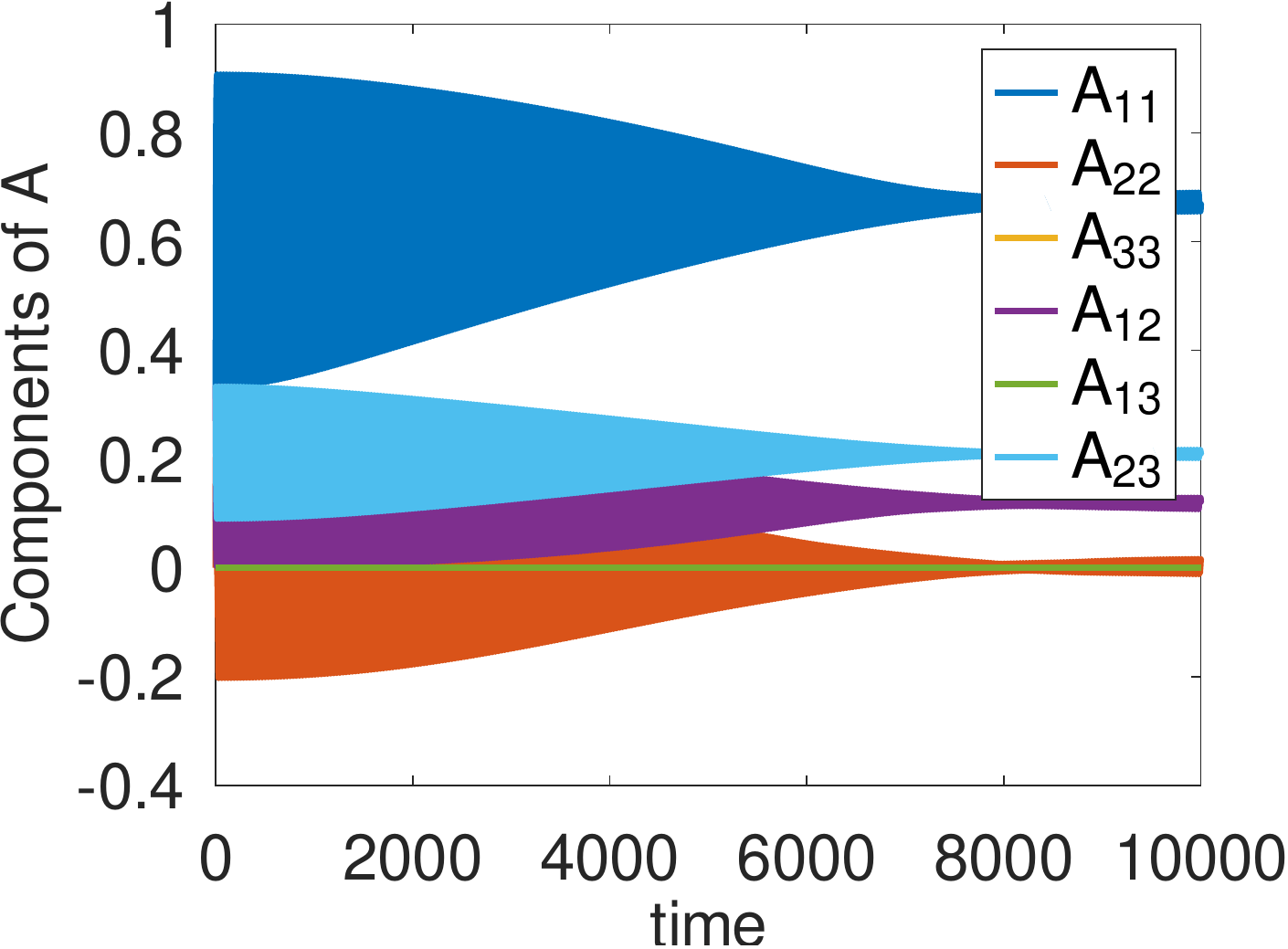}
\end{center}
\caption{Shear stress and averaged over $y$ of components of $\mathsf A$ for shear flow with two dimensional fiber orientations perturbed by $\epsilon = 0.01$.}
\label{fig2b-verylong}
\end{figure}

\begin{figure}
\begin{center}
\includegraphics[scale=0.4]{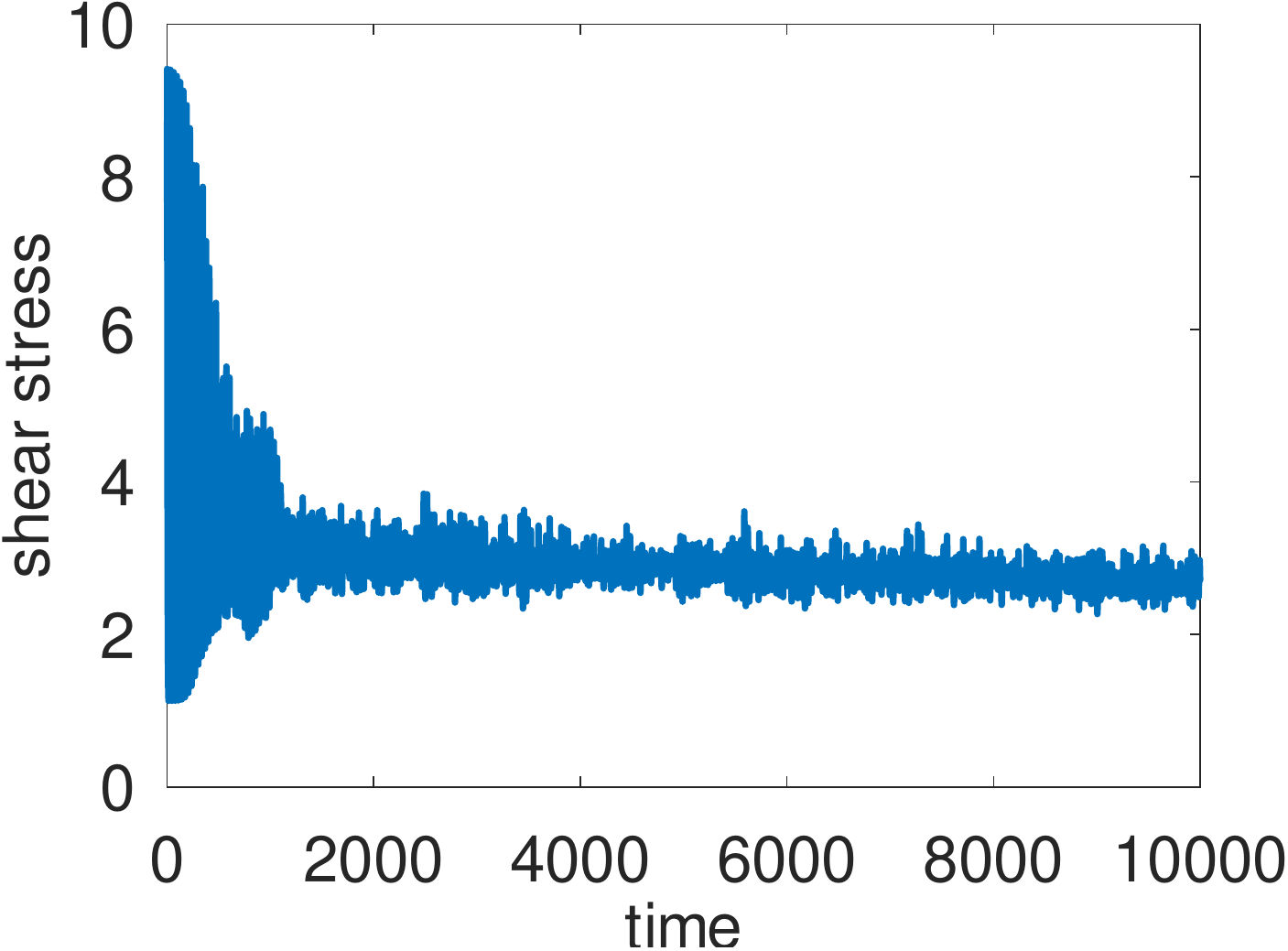}
\includegraphics[scale=0.4]{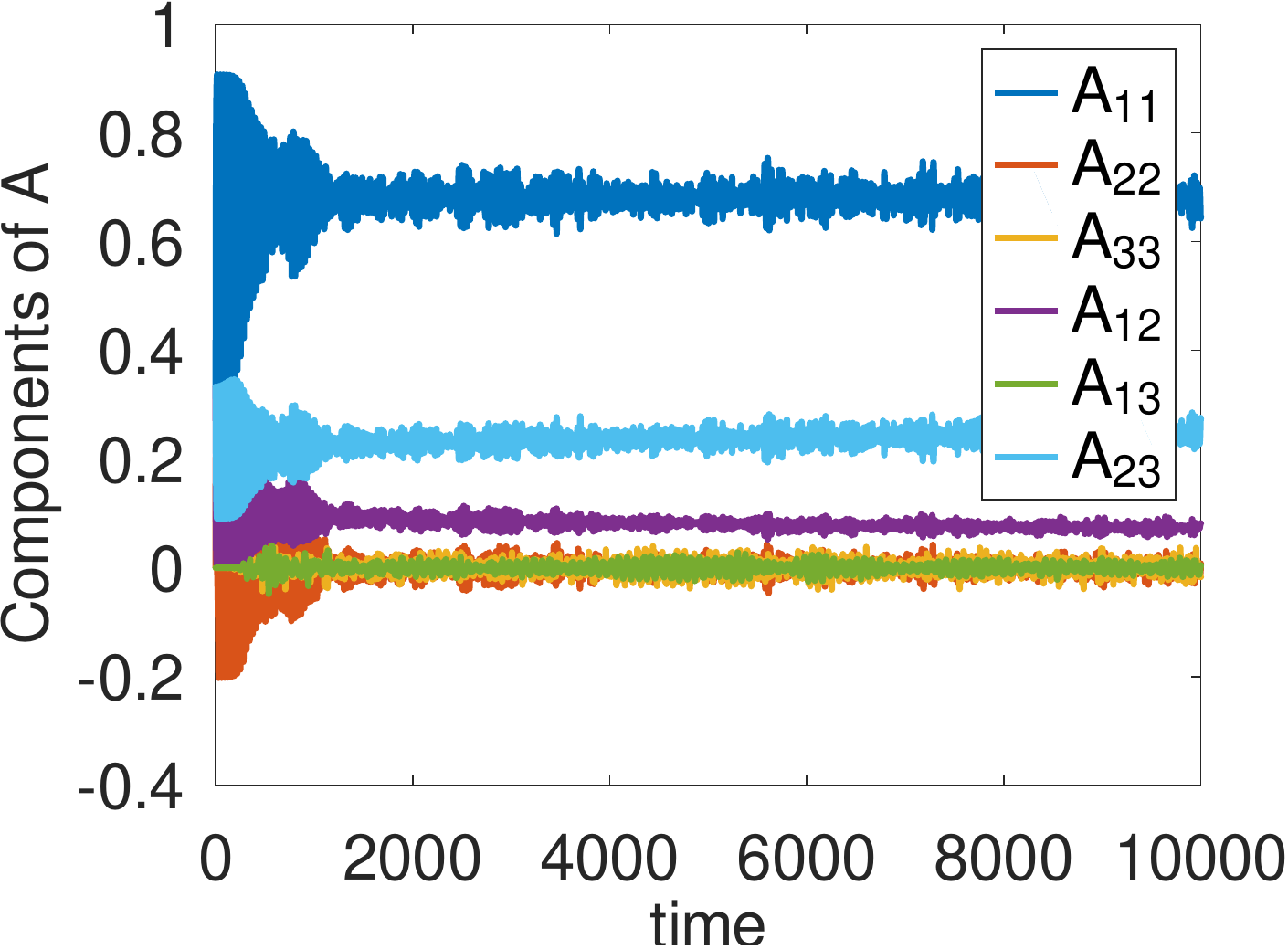}
\end{center}
\caption{Shear stress and averaged over $y$ of components of $\mathsf A$ for shear flow with unrestricted fiber orientations perturbed by $\epsilon = 0.01$.}
\label{fig3b-verylong}
\end{figure}

\begin{table}
\begin{tabular}{|c|c|c|c|c|}
\hline
& mean for & S.D. for & mean for & S.D. for \\
& $\epsilon = 0.01$ & $\epsilon = 0.01$ & $\epsilon = 0.3$ & $\epsilon = 0.3$ \\
\hline
$\Sigma_1$ & 3.185 & 0.896 & 2.763 & 0.177 \\
$\mathsf A_{11}$ & 0.696 & 0.073 & 0.682 & 0.02 \\
$\mathsf A_{22}$ & 0 & 0.055 & 0 & 0.012 \\
$\mathsf A_{33}$ & 0 & 0.014 & 0 & 0.012 \\
$\mathsf A_{12}$ & 0.09 & 0.036 & 0.075 & 0.007 \\
$\mathsf A_{13}$ & -0.001 & 0.014 & 0 & 0.008 \\
$\mathsf A_{23}$ & 0.213 & 0.037 & 0.242 & 0.014 \\
\hline
\end{tabular}

\

\caption{Mean and standard deviations of shear stress, and the averaged over $y$ of the components of $\mathsf A$, for unrestricted perturbations over the second half of the respective time interval.}
\label{steady-state}
\end{table}

\begin{figure}
\begin{center}
\includegraphics[scale=0.4]{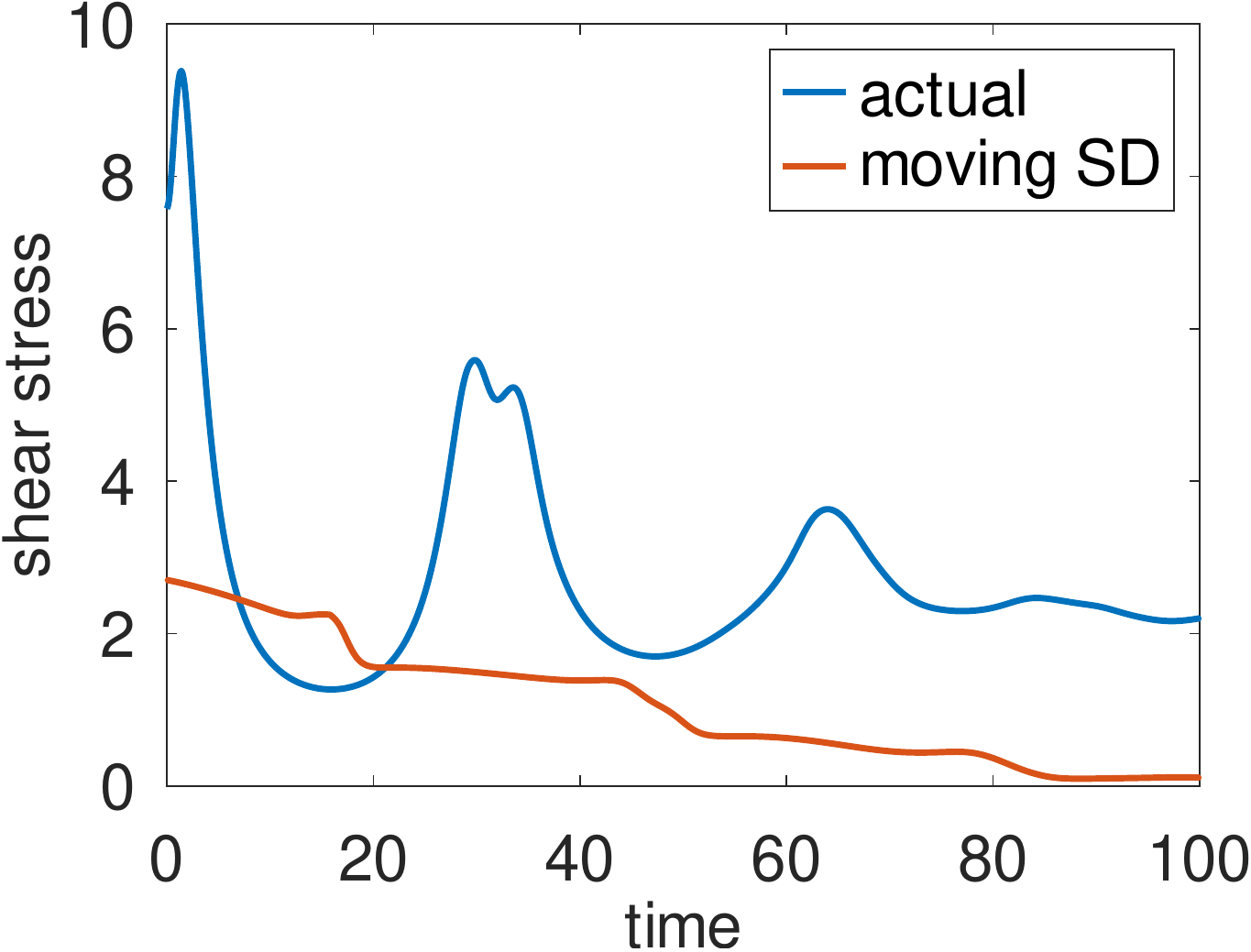}
\end{center}
\caption{An example of a moving standard deviation of the shear stress.}
\label{moving sd}
\end{figure}

\begin{figure}
\begin{center}
\includegraphics[scale=0.4]{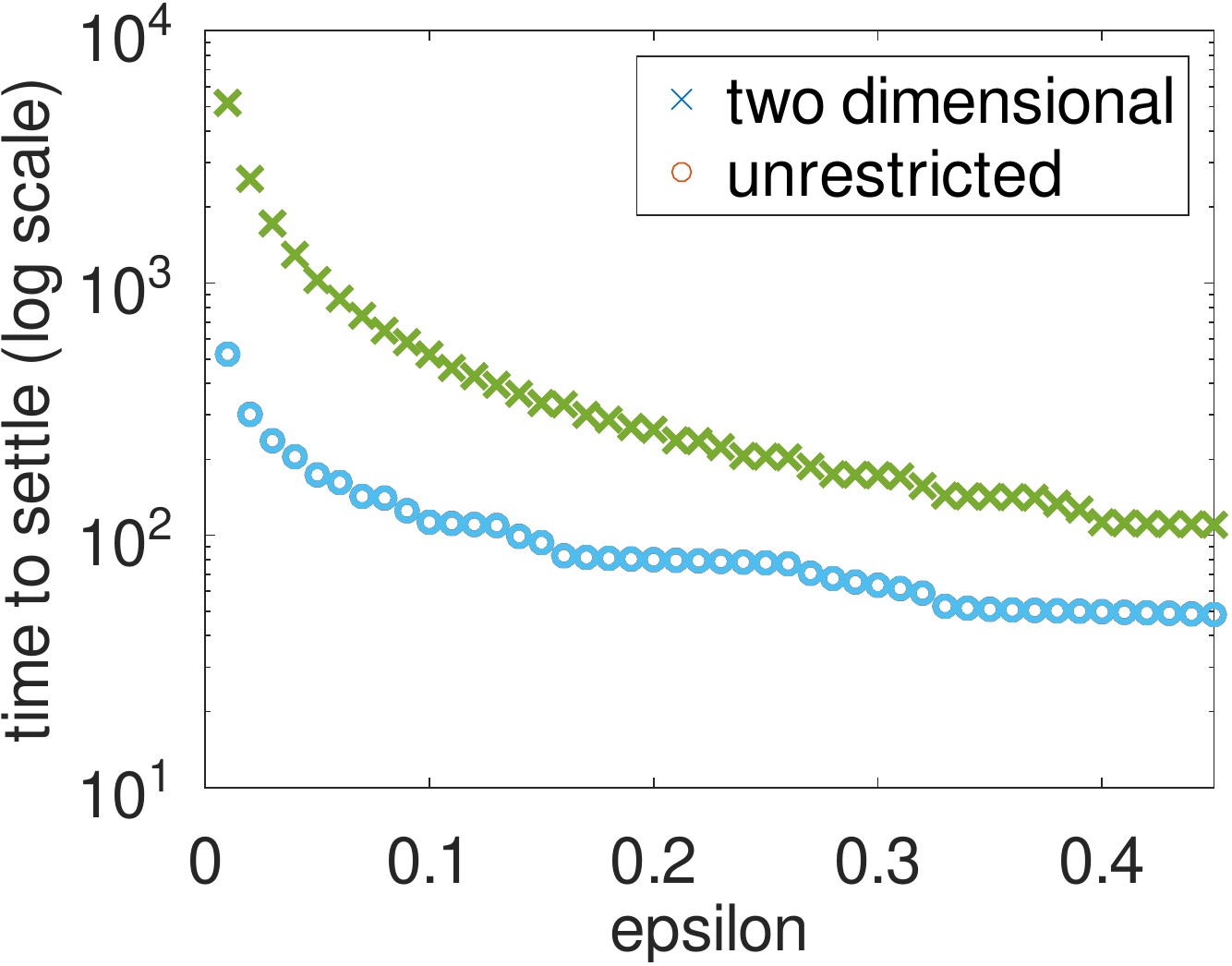}
\end{center}
\caption{Times to settle versus $\epsilon$, for two dimensional perturbations and unrestricted perturbations, with $\lambda = 0.98$.}
\label{time-to-settle-eps}
\end{figure}

\begin{figure}
\begin{center}
\includegraphics[scale=0.4]{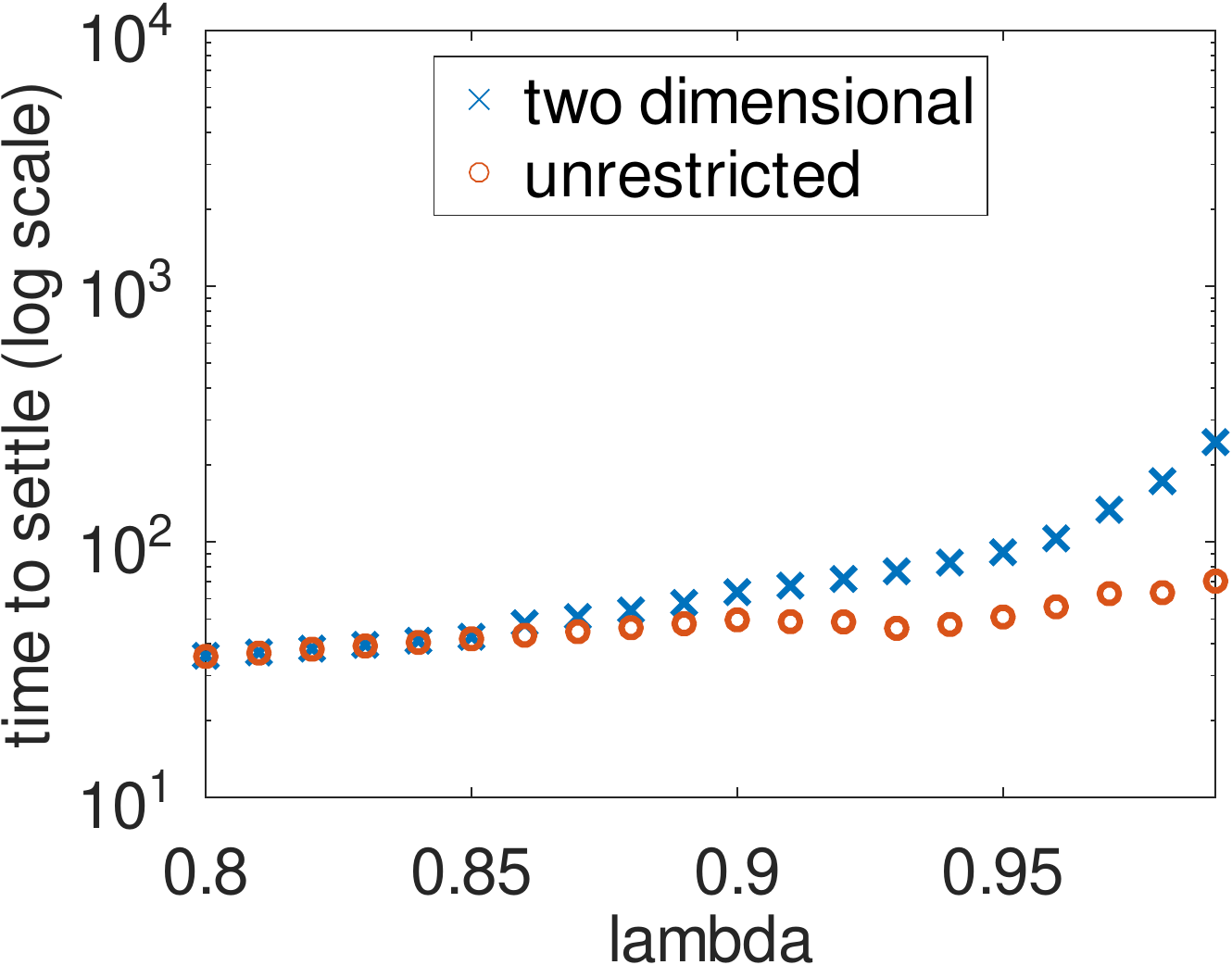}
\includegraphics[scale=0.4]{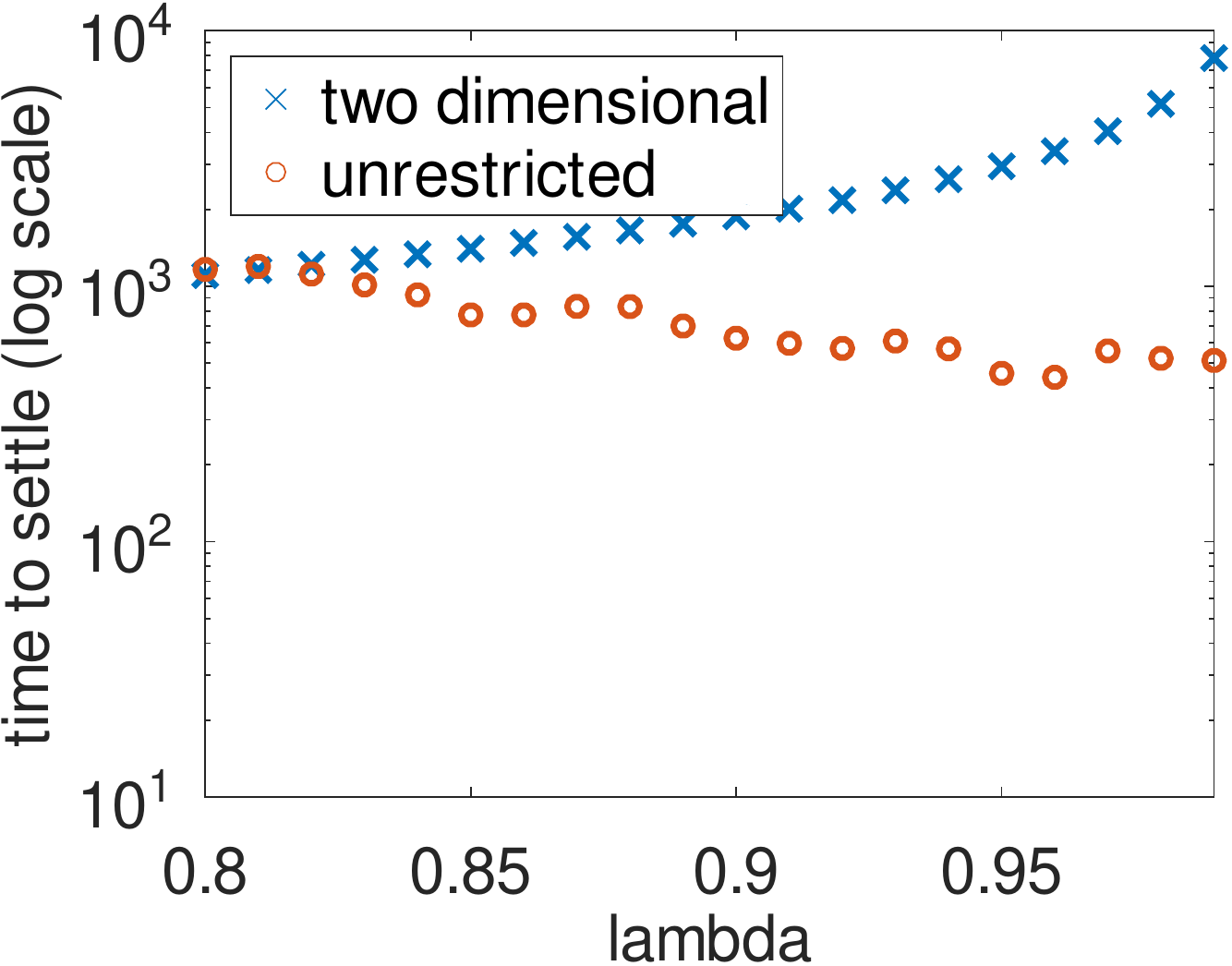}
\end{center}
\caption{Times to settle versus $\lambda$, for two dimensional perturbations and unrestricted perturbations, with $\epsilon = 0.3$ (left hand side) and $\epsilon = 0.01$ (right hand side).}
\label{time-to-settle-lambda}
\end{figure}

\section{Acknowledgments}

The author gratefully acknowledges support from N.S.F.\ grant C.M.M.I.\ 0727399.

\bibliographystyle{plain}

\begin{thebibliography}{10}

\bibitem{anczurowski:67}
E.~Anczurowski and S.G. Mason.
\newblock The kinetics of flowing dispertions iii. equilibrium orientation of
  rods and discs (experimental).
\newblock {\em J. Colloid Int. Sci.}, 23:533--546, 1967.

\bibitem{batchelor:71}
G.K. Batchelor.
\newblock Stress generated in a non-dilute suspension of elongated particles by
  pure straining motion.
\newblock {\em Journal of Fluid Mechanics}, 46:813--829, 1971.

\bibitem{bird:87b}
R.B. Bird, C.F. Curtiss, R.~C. Armstrong, and O.~Hassager.
\newblock {\em {Dynamics of Polymeric Liquids}}, volume 2: Kinetic Theory.
\newblock John Wiley \& Sons, Inc., New York, NY, 2nd edition, 1987.

\bibitem{chicone:06}
C.~Chicone.
\newblock {\em Ordinary Differential Equations with Applications}.
\newblock Springer-Verlag, New York, 2nd edition, 2006.

\bibitem{dinh:84}
S.M. Dinh and R.C. Armstrong.
\newblock {A Rheological Equation of State for Semiconcentrated Fiber
  Suspensions}.
\newblock {\em Jn. of Rheology}, {\bf 28}(3):207--227, 1984.

\bibitem{folgar:84}
F.P. Folgar and C.L. Tucker.
\newblock {Orientation Behavior of Fibers in Concentrated Suspensions}.
\newblock {\em Jn. of Reinforced Plastics and Composites}, {\bf 3}:98--119,
  April 1984.

\bibitem{jeffery:23}
G.B. Jeffery.
\newblock {The Motion of Ellipsoidal Particles Immersed in a Viscous Fluid}.
\newblock {\em {Proceedings of the Royal Society of London A}}, {\bf
  102}:161--179, March 1922.

\bibitem{lipscomb:88}
G.G. Lipscomb~II, M.M. Denn, D.U. Hur, and D.V. Boger.
\newblock {Flow of Fiber Suspensions in Complex Geometries}.
\newblock {\em {Jn. of Non-Newtonian Fluid Mechanics}}, {\bf 26}:297--325,
  1988.

\bibitem{montgomery-smith:10d}
S.J. Montgomery-Smith.
\newblock {Perturbations of the coupled Jeffery-Stokes equations}.
\newblock {\em J. of Fluid Mechanics}, 681:622--638, 2011.

\bibitem{montgomery-smith:10d-corrigendum}
S.J. Montgomery-Smith.
\newblock {Perturbations of the coupled Jeffery-Stokes equations ---
  Corrigendum}, 2011.
\newblock
  \url{https://stephenmontgomerysmith.github.io//preprints/jeff-stokes-corrigendum.pdf}.

\bibitem{montgomery:10b}
S.J. Montgomery-Smith, Wei He, D.A. Jack, and D.E. Smith.
\newblock {Exact Tensor Closures for the Three Dimensional Jeffery's Equation}.
\newblock {\em {J. of Fluid Mechanics}}, {\bf 680}:321--335, 2011.

\bibitem{sepehr:04}
M.~Sepehr, P.J. Carreau, M.~Grmela, G.~Ausias, and P.G. Lafleur.
\newblock {Comparison of Rheological Properties of Fiber Suspensions with Model
  Predictions}.
\newblock {\em Jn. of Polymer Engineering}, {\bf 24}(6):579--610, 2004.

\bibitem{shaqfeh:90a}
E.S.G. Shaqfeh and G.H Fredrickson.
\newblock The hydrodynamic stress in a suspension.
\newblock {\em Physics of Fluids A}, 2:7--24, 1990.

\bibitem{szeri:96}
A.J. Szeri and D.J. Lin.
\newblock A deformation tensor model of brownian suspensions of orientable
  particles ---the nonlinear dynamics of closure models.
\newblock {\em Journal of Non-Newtonian Fluid Mechanics}, 64:43--69, 1996.

\bibitem{verleye:93}
V.~Verleye and F.~Dupret.
\newblock {Prediction of Fiber Orientation in Complex Injection Molded Parts}.
\newblock In {\em {Developments in Non-Newtonian Flows}}, pages 139--163, 1993.

\bibitem{verweyst:98}
B.E. VerWeyst.
\newblock {\em {Numerical Predictions of Flow Induced Fiber Orientation in
  Three-Dimensional Geometries}}.
\newblock PhD thesis, {University of Illinois at Urbana Champaign}, 1998.

\bibitem{wang:08}
J.~Wang, J.F.\ O'Gara, and C.L. Tucker.
\newblock An objective model for slow orientation kinetics in concentrated
  fiber suspensions: Theory and rheological evidence.
\newblock {\em J. Rheology}, {\bf 52}:1179--1200, 2008.

\end{thebibliography}

\end{document}